\documentclass[journal,11pt,draftclsnofoot,onecolumn]{IEEEtran}

\IEEEoverridecommandlockouts                              

\usepackage{graphicx}
\usepackage[square,comma,numbers,sort&compress]{natbib}
\usepackage{amsthm,amssymb,amscd,amssymb,amsmath,mathrsfs}
\usepackage{epstopdf}
\usepackage{balance}
\usepackage[noend]{algorithm} 
\usepackage{algorithmicx} 
\usepackage[noend]{algpseudocode}
\usepackage{multirow} 
\usepackage{amsmath}
\usepackage{xcolor}
\usepackage{lineno}

\makeatletter

\newcommand{\Rmnum}[1]{\expandafter\@slowromancap\romannumeral #1@}
\makeatother

\theoremstyle{plain}
\newtheorem{theorem}{Theorem}

\newtheorem{lemma}{Lemma}

\theoremstyle{definition}
\newtheorem{definition}{Definition}

\theoremstyle{remark}

\begin{document}
%
\title{Supervisor Localization of Discrete-Event Systems under Partial Observation*
 }

\author{Renyuan Zhang$^{1}$, Kai Cai$^{2}$, W.M. Wonham$^{3}$
\thanks{*This work was supported in part by the National Nature Science Foundation of China, Grant no. 61403308;
the Program to Disseminate Tenure Tracking System, MEXT, Japan;
the Natural Sciences and Engineering Research Council, Canada, Grant no. 7399.
}
\thanks{$^{1}$R. Zhang is with School of Automation, Northwestern Polytechnical University, China
        {\tt\small ryzhang@nwpu.edu.cn}}%
\thanks{$^{2}$K. Cai is with Urban Research Plaza, Osaka City University, Japan
        {\tt\small kai.cai@eng.osaka-cu.ac.jp}}%
\thanks{$^{3}$W.M. Wonham is with the Systems Control Group, Department of Electrical and
        Computer Engineering, University of Toronto, Canada
        {\tt\small wonham@ece.utoronto.ca.}}%
}

\maketitle

\thispagestyle{empty} \pagestyle{plain}


\begin{abstract}
Recently we developed \emph{supervisor localization}, a top-down
approach to distributed control of discrete-event systems. Its
essence is the allocation of monolithic (global) control action
among the local control strategies of individual agents. In this
paper, we extend supervisor localization by considering partial
observation; namely not all events are observable. Specifically, we
employ the recently proposed concept of \emph{relative
observability} to compute a partial-observation
monolithic supervisor, and then design a suitable localization
procedure to decompose the supervisor into a set of local
controllers. In the resulting local controllers, only observable
events can cause state change. Further, to deal with large-scale
systems, we combine the partial-observation supervisor localization
with an efficient architectural synthesis approach: first compute a
heterarchical array of partial-observation
decentralized supervisors and coordinators, and then localize each
of these supervisors/coordinators into local controllers.

\end{abstract}


\section{Introduction}


In \cite{CaiWonham:2010a,CaiWonham:2010b,ZhangCai:2013,CaiWonham:2015,CaiWon:book15} we developed a
top-down approach, called {\it supervisor localization}, to the
distributed control of multi-agent discrete-event systems (DES).
This approach first synthesizes a monolithic supervisor (or a
heterarchical array of modular supervisors) assuming that {\it all} events
can be observed, and then decomposes the supervisor into a set of
local controllers for the component agents. Localization
creates a purely distributed control architecture in which each
agent is controlled by its own local controller; this is particularly
suitable for applications consisting of many autonomous components,
e.g. multi-robot systems. Moreover, localization can significantly
improve the comprehensibility of control logic, because the resulting local
controllers typically have many fewer states than their parent supervisor.
The assumption of full
event observation, however, may be too strong in practice, since
there often lacks enough sensors to observe every event.

In this paper, we extend supervisor localization to address the
issue of partial observation. Our approach is as follows.
We first
synthesize a partial-observation monolithic
supervisor using the concept of {\it relative observability} in
\cite{CaiZhangWonham:2015}. Relative observability is
generally stronger than observability \cite{LinWonham:1988,Cieslak:1988}, weaker than normality
\cite{LinWonham:1988,Cieslak:1988}, and the supremal relatively observable (and controllable)
sublanguage of a given language exists. The supremal
sublanguage may be effectively computed \cite{CaiZhangWonham:2015}, and then implemented by a
partial-observation (feasible and nonblocking) supervisor
\cite[Chapter~6]{Wonham:2015a}. We then suitably extend the
localization procedure in \cite{CaiWonham:2010a} to decompose the
supervisor into local controllers for individual agents, and
moreover prove that the derived local controlled behavior is
equivalent to the monolithic one. We then suitably extend the
localization procedure in \cite{CaiWonham:2010a} to decompose the
supervisor into local controllers for individual agents, and
moreover prove that the derived local controlled behavior is
equivalent to the monolithic one.

The main contribution of this work is the novel combination of
supervisor localization \cite{CaiWonham:2010a} with relative
observability \cite{CaiZhangWonham:2015}, which leads to a
systematic approach to distributed control of DES under partial
observation. The central concept of supervisor localization is {\it
control cover} \cite{CaiWonham:2010a}, which is defined on the state
set of the full-observation supervisor. Under partial observation,
we propose an extended control cover, which is defined on the state
set of the partial-observation supervisor; roughly speaking, the
latter corresponds to the \emph{powerset} of the full-observation
supervisor's state set. In this way, in the transition structure of
the resulting local controllers, only observable events can lead to
state changes. We design an extended localization algorithm for
computing these local controllers. Moreover, to deal with
large-scale systems, we combine the developed localization procedure
with an efficient architectural synthesis approach
\cite{FengWonham:2008}: first compute a heterarchical array of
partial-observation decentralized supervisors and coordinators that
collectively achieves globally feasible and nonblocking controlled
behavior, and then localize each of these supervisors/coordinators
into local controllers.

Our proposed localization procedure can in principle be used to
construct local controllers from a partial-observation
supervisor computed by any synthesis method. In particular, the
algorithms in \cite{TakaiUshio:2003,YinLafortune:2016} compute a nonblocking
(maximally) observable sublanguage that is generally incomparable
with the supremal relatively observable sublanguage. The reason that
we adopt relative observability is first of all that its
generator-based computation of the supremal sublanguage is better
suited for applying our localization algorithm; by contrast
\cite{YinLafortune:2016} uses a different transition structure called
``bipartite transition system''. Another important reason is that
the computation of relative observability has been implemented and
tested on a set of benchmark examples. This enables us to study
distributed control under partial observation of more realistic
systems; by contrast, the examples reported in
\cite{TakaiUshio:2003,YinLafortune:2016} are limited to academic ones.

We note that in \cite{SampathLafortune:1998,DubreilEt:2010,YinLafortune:2016a}
partial-observation supervisors are synthesized to enforce properties other
than safety (satisfying imposed specifications) and nonblockingness;
examples include diagnosability and opacity. Although we focus on safety
and nonblockingness, for the sake of consistency with our previous work
on full-observation localization that targets multi-agent distributed control
problems, our developed localization procedure may be applied in the same way to
decompose partial-observation supervisors with other properties. Thus if a
partial-observation supervisor enforces specified properties determined only
by the synthesized language, then those
properties are preserved by localization and achieved collectively by the
synthesized local controllers.

The paper is organized as follows. Section II reviews the supervisory
control problem of DES under partial observation and formulates the
partial-observation supervisor localization problem. Section III
develops the partial-observation localization procedure, and Section
IV presents the localization algorithm, which is illustrated by a
Transfer Line example. Section V outlines a procedure combining the
partial-observation localization with an efficient heterarchical
supervisor synthesis to address distributed control of large-scale
systems.
Finally Section VI states our conclusions.


\section{Preliminaries and Problem Formulation}

\subsection{Supervisory Control of DES under Partial Observation}


A DES plant is given by a generator
\begin{align} \label{eq:plant}
{\bf G} = (Q,\Sigma, \delta, q_0, Q_m)
\end{align}
where $Q$ is the finite state set; $q_0 \in Q$ is the initial state;
$Q_m \subseteq Q$ is the subset of marker states; $\Sigma$ is the finite event set;
$\delta: Q\times \Sigma\rightarrow Q$ is the (partial) state transition function.
In the usual way, $\delta$ is extended to $\delta:Q\times\Sigma^*\rightarrow Q$,
and we write $\delta(q,s)!$ to mean that $\delta(q,s)$ is defined.
Let $\Sigma^*$ be the set of all finite strings, including the empty string $\epsilon$.
The {\it closed behavior} of $\bf G$ is the language
\[L({\bf G}) = \{s\in \Sigma^*|\delta(q_0,s)!\}\]
and the {\it marked behavior} is
\[L_m({\bf G}) = \{s\in L({\bf G})|\delta(q_0,s)\in Q_m\}\subseteq L({\bf G}).\]
A string $s_1$ is a {\it prefix} of a string $s$, written $s_1\leq
s$, if there exists $s_2$ such that $s_1s_2 = s$. The {\it (prefix)
closure} of $L_m({\bf G})$ is $\overline{L_m({\bf G})} := \{s_1 \in
\Sigma^*|(\exists s\in L_m({\bf G}))~s_1\leq s\}$. In this paper, we
assume that $\overline{L_m({\bf G})} = L({\bf G})$; namely, $\bf G$
is {\it nonblocking}.

For supervisory control, the event set $\Sigma$ is partitioned into
$\Sigma_c$, the subset of controllable events that can be disabled
by an external supervisor, and $\Sigma_{uc}$, the subset of
uncontrollable events that cannot be prevented from occurring (i.e.
$\Sigma = \Sigma_c\dot\cup\Sigma_{uc}$). For partial observation,
$\Sigma$ is partitioned into $\Sigma_o$, the subset of observable
events, and $\Sigma_{uo}$, the subset of unobservable events (i.e.
$\Sigma = \Sigma_o \dot\cup \Sigma_{uo}$). Bring in the
\emph{natural projection} $P : \Sigma^* \rightarrow \Sigma_o^*$
defined by
\begin{equation} \label{eq:natpro}
\begin{split}
P(\epsilon) &= \epsilon; \\
P(\sigma) &= \left\{
  \begin{array}{ll}
    \epsilon, & \hbox{if $\sigma \notin \Sigma_o$,} \\
    \sigma, & \hbox{if $\sigma \in \Sigma_o$;}
  \end{array}
\right.\\
P(s\sigma) &= P(s)P(\sigma),\ \ s \in \Sigma^*, \sigma \in \Sigma.
\end{split}
\end{equation}
As usual, $P$ is extended to $P : Pwr(\Sigma^*) \rightarrow Pwr(\Sigma_o^*)$, where
$Pwr(\cdot)$ denotes powerset. Write $P^{-1}: Pwr(\Sigma_o^*) \rightarrow Pwr(\Sigma^*)$
for the \emph{inverse-image function} of $P$.

For two languages $L_1 \subseteq \Sigma_1^*$ and $L_2 \subseteq \Sigma_2^*$,
the {\it synchronous product} $L_1 || L_2 \subseteq (\Sigma_1 \cup \Sigma_2)^*$
is defined according to $L_1 || L_2 := P_1^{-1}L_1 \cap P_2^{-1}L_2$,
where $P_i:(\Sigma_1\cup\Sigma_2)^*\rightarrow \Sigma_i^*$ ($i = 1, 2$)
are the natural projections as defined in (\ref{eq:natpro}). For two
generators ${\bf G}_i = (Q_i, \Sigma_i, \delta_i, q_{0,i}, Q_{m,i})$, $i = 1, 2$, 
let $L_m({\bf G}_i)$ and $L({\bf G}_i)$ be the marked and
closed behaviors of ${\bf G}_i$ respectively; then the
{\it synchronous product} ${\bf G} = (Q,\Sigma,\delta,q_0,Q_m)$ of ${\bf G}_1$ and
${\bf G}_2$, denoted by ${\bf G}_1 || {\bf G}_2$,
is constructed \cite{Wonham:2015a} to have marked behavior $L_m({\bf G}) = L_m({\bf G}_1) || L_m({\bf G}_2)$ and
closed behavior $L({\bf G}) = L({\bf G}_1) || L({\bf G}_2)$.
Synchronous product of more than two generators can be constructed similarly.


A {\it supervisory control} for $\bf G$ is any map $V:L({\bf G})\rightarrow \Gamma$,
where $\Gamma := \{\gamma \subseteq \Sigma|\gamma \supseteq \Sigma_{uc}\}$. Then the
closed-loop system is $V/{\bf G}$, with closed behavior $L(V/{\bf G})$ and marked
behavior $L_m(V/{\bf G})$ \cite{Wonham:2015a}.
Under partial observation $P:\Sigma^*\rightarrow \Sigma_o^*$,
we say that $V$ is {\it feasible} if
\begin{equation}\label{eq:feaisble}
(\forall s, s' \in L({\bf G}))~ P(s) = P(s')\Rightarrow V(s) = V(s'),
\end{equation}

and $V$ is {\it nonblocking} if $\overline{L_m(V/{\bf G})} = L(V/{\bf G})$.

It is well-known \cite{LinWonham:1988} that under partial observation, a feasible
and nonblocking supervisory control $V$ exists which synthesizes a (nonempty)
sublanguage $K\subseteq L_m({\bf G})$ if and only if $K$ is both
controllable and observable \cite{Wonham:2015a}. When $K$ is not observable,
however, there generally does not exist the supremal observable (and
controllable) sublanguage of $K$. Recently in \cite{CaiZhangWonham:2015},
a new concept of {\it relative observability} is proposed, which is stronger
than observability but permits the existence of the supremal
relatively observable sublanguage.

Formally, a sublanguage $K \subseteq L_m({\bf G})$ is {\it
controllable} \cite{Wonham:2015a} if
\begin{equation} \label{eq:defcontrol}
\overline{K}\Sigma_{uc}\cap L({\bf G}) \subseteq \overline{K}.
\end{equation}
Let $C\subseteq L_m({\bf G})$. A sublanguage $K \subseteq C$ is {\it relatively observable} with
respect to $C$ (or $C$-observable) if for every
pair of strings $s,s'\in \Sigma^*$ that are lookalike under $P$,
i.e. $P(s)=P(s')$, the following two conditions hold
\cite{CaiZhangWonham:2015}:
\begin{align}
\mbox{(i)}~ &(\forall \sigma \in \Sigma) s\sigma \in \overline{K}, s'
\in \overline{C},s'\sigma\in L({\bf G})
\Rightarrow s'\sigma \in \overline{K} \label{eq:sub1:reloberv} \\
\mbox{(ii)}~ &s \in K, s' \in \overline{C} \cap L_m({\bf G})
\Rightarrow s' \in K \label{eq:sub2:reloberv}
\end{align}
For $E\subseteq L_m({\bf G})$ write $\mathcal{CO}(E)$ for the family of
controllable and $C$-observable sublanguages of $E$. Then
$\mathcal{CO}(E)$ is nonempty (the empty language $\emptyset$
belongs) and is closed under set union; $\mathcal{CO}(E)$ has a
unique supremal element $\sup \mathcal{CO}(E)$ given by
\begin{equation*}
\sup \mathcal{CO}(E) = \bigcup\{K|K\in \mathcal{CO}(E)\}
\end{equation*}
which may be effectively computed \cite{CaiZhangWonham:2015}. Note that since relative
observability is weaker than normality \cite{Wonham:2015a}, $\sup \mathcal{CO}(E)$
is generally larger than the normality counterpart.




\subsection{Formulation of Partial-Observation Localization Problem}


Let the plant $\textbf{G}$ be comprised of $N$ ($>1$)
component agents
\begin{equation*}
{\bf G}_k = (Q_k, \Sigma_k, \delta_k, q_{0,k}, Q_{m,k}),\ \ \ k =
1,...,N.
\end{equation*}
Then ${\bf G}$ is the synchronous product of  ${\bf G}_k$, $k$ in
the integer range $\{1,...,N\}$, denoted as $[1,N]$, i.e. ${\bf G} = ||
\{{\bf G}_k|k \in [1,N]\}$. Here, the $\Sigma_k$ need not be pairwise
disjoint. These agents are implicitly coupled through a
specification language $E \subseteq \Sigma^*$ that imposes a
constraint on the global behavior of $\textbf{G}$ ($E$ may itself be
the synchronous product of multiple component specifications).
For the plant $\textbf{G}$ and the imposed
specification $E$, let the generator $\textbf{SUP} = (X, \Sigma,
\xi, x_0, X_m)$ be such that
\begin{equation} \label{eq:monosup}
L_m(\textbf{SUP}) := \sup \mathcal {CO}(E \cap L_m(\textbf{G})).
\end{equation}
and $L({\bf SUP}) = \overline{L_m({\bf SUP})}$ (i.e. $\bf SUP$ is
nonblocking).
We call {\bf SUP} the \emph{controllable and observable controlled
behavior}.\footnote{Note that {\bf SUP}, defined over the entire
event set $\Sigma$, is \emph{not} a representation of a
partial-observation supervisor. The latter can only have observable
events as state transitions, according to the definition in
Section~\ref{subsec:uncertaintyset}, below.} To rule out the trivial
case, we assume that $L_m(\textbf{SUP}) \neq \emptyset$.


Now let $\alpha \in \Sigma_c$ be an arbitrary controllable event,
which may or may not be observable. We say that a generator
\begin{equation*}
{\bf LOC}_\alpha =
(Y_\alpha,\Sigma_\alpha,\eta_\alpha,y_{0,\alpha},Y_{m,\alpha}),\
\Sigma_\alpha \subseteq \Sigma_o \cup \{\alpha\}
\end{equation*}
is a {\it partial-observation local controller} for $\alpha$ if (i)
${\bf LOC}_\alpha$ enables/disables the event $\alpha$ (and only
$\alpha$) consistently with ${\bf SUP}$, and (ii) if $\alpha$ is
unobservable, then $\alpha$-transitions are selfloops in ${\bf
LOC}_\alpha$, i.e.
\begin{align*}
(\forall y \in Y_\alpha)\ \eta_\alpha(y,\alpha)! \Rightarrow
\eta_\alpha(y,\alpha) = y.
\end{align*}

Condition~(i) means that for all $s \in \Sigma^*$ there holds
\begin{align} \label{eq:loc1}
P_\alpha(s)\alpha \in L({{\bf LOC}_\alpha}),\ s\alpha \in L({\bf G}),~
&s \in L({\bf SUP})\notag\\
\Leftrightarrow &s\alpha \in L({\bf SUP})
\end{align}
where $P_\alpha:\Sigma^*\rightarrow \Sigma_\alpha^*$ is the natural
projection. Condition~(ii) requires that only observable events may
cause a state change in ${\bf LOC}_\alpha$, i.e.
\begin{align} \label{eq:loc2}
(\forall y, y' \in Y_\alpha, \forall \sigma \in \Sigma_\alpha)\ y' =
\eta_\alpha(y,\sigma)!,\ y' \neq y \Rightarrow \sigma \in \Sigma_o.
\end{align}
This requirement is a distinguishing feature of a partial-observation
local controller as compared to its full-observation counterpart in
\cite{CaiWonham:2010a}.

Note that the event set $\Sigma_\alpha$ of ${\bf LOC}_\alpha$ in
general satisfies
\begin{equation*}
\{\alpha\} \subseteq \Sigma_\alpha \subseteq \Sigma_o \cup
\{\alpha\};
\end{equation*}
in typical cases, both subset containments are strict. The
events in $\Sigma_\alpha \setminus \{ \alpha \}$ may be viewed as
communication events that are critical to achieve synchronization with other
partial-observation local controllers (for other controllable
events). The event set $\Sigma_\alpha$ is not fixed \emph{a priori}, but will
be determined as part of the localization result presented in the
next section.

We now formulate the {\it Partial-Observation Supervisor Localization
Problem}:

Construct a set of partial-observation local controllers $\{{\bf
LOC}_\alpha\ |\ \alpha \in \Sigma_c\}$ such that the collective
controlled behavior of these local controllers is equivalent to the
controllable and observable controlled behavior ${\bf SUP}$ in
(\ref{eq:monosup}) with respect to $\bf G$, i.e.
\begin{align*}
   L_m({\bf G})\cap \Big(\mathop \bigcap\limits_{\alpha \in \Sigma_{c}}
   P_\alpha^{-1}L_m({\bf LOC}_{\alpha}) \Big) &= L_m({\bf SUP}) \\
   L({\bf G}) \cap \Big(\mathop \bigcap\limits_{\alpha \in \Sigma_{c}}
   P_\alpha^{-1}L({\bf LOC}_{\alpha}) \Big)  &= L({\bf SUP}).
\end{align*}

Having obtained a set of partial-observation local controllers, one
for each controllable event, we can allocate each controller to the
agent(s) owning the corresponding controllable event. Thereby we
build for a multi-agent DES a nonblocking distributed control
architecture under partial observation.


\section{Partial-Observation Localization Procedure}

\subsection{Uncertainty Set} \label{subsec:uncertaintyset}

Let ${\bf G} = (Q,\Sigma,\delta,q_0,Q_m)$ be the plant,
$\Sigma_o \subseteq \Sigma$ the subset of observable events, and $P:
\Sigma^* \rightarrow \Sigma^*_o$ the corresponding natural
projection.
Also let ${\bf SUP} = (X,\Sigma,\xi,x_0,X_m)$ be the
controllable and observable controlled behavior (as
defined in (\ref{eq:monosup})).

Under partial observation, when a string $s \in L({\bf SUP})$
occurs, what is observed is $P(s)$; namely, the events in
$\Sigma_{uo}$ $(=\Sigma \setminus \Sigma_o)$ are erased. Hence two
different strings $s$ and $s'$ may be lookalike, i.e. $P(s) =
P(s')$.
For $s \in L({\bf SUP})$, let $U(s)$ be the subset of
states that may be reached by some string $s'$ that looks like $s$,
i.e.
\begin{equation*}
U(s) = \{x\in X|(\exists s' \in \Sigma^*) P(s) = P(s'), x =
\xi(x_0,s')\}.
\end{equation*}
It is always true that the state $\xi(x_0,s) \in U(s)$. We call
$U(s)$ the {\it uncertainty set} of the state $\xi(x_0,s)$ associated with string $s$.
Let
\begin{equation} \label{eq:uncertainset}
\mathcal{U}(X) := \{U(s) \subseteq X | s\in L({\bf SUP})\}
\end{equation}
i.e. $\mathcal{U}(X)$ is the set of uncertainty sets of all states
(associated with strings in $L({\bf SUP})$) in $X$. The size of $\mathcal{U}(X)$ is $|\mathcal{U}(X)| \leq
2^{|X|}$ in general.

\begin{figure}[!t]
\centering
    \includegraphics[scale = 0.2]{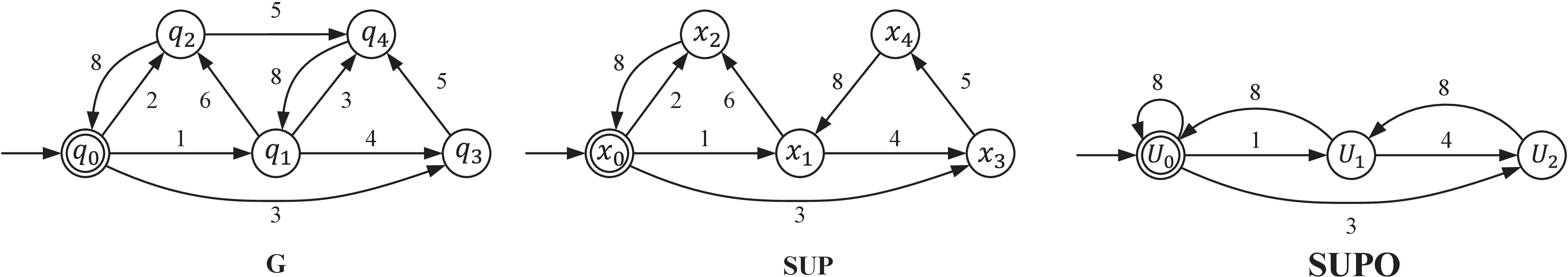}
\caption{Plant $\bf G$, controllable and observable
controlled behavior $\bf SUP$, partial-observation monolithic
supervisor ${\bf SUPO}$, $\Sigma_c = \{1,3,5\}$, and $\Sigma_o =
\{1,3,4,8\}$. Inspecting the transition diagram of $\bf SUP$,
the uncertainty sets are $U(\epsilon) = \{x_0,x_2\}$, $U(1) =
\{x_1,x_2\}$, $U(1.4) = \{x_3,x_4\}$, and for the remainder of
strings $s \in L({\bf SUP})$, $U(s)$ equals one of the above;
namely, the set $\mathcal{U}(X)$ of uncertainty sets of $\bf SUP$ is
$\mathcal{U}(X) = \{\{x_0,x_2\},\{x_1,x_2\},\{x_3,x_4\}\}$. For
later reference, denote $U_0 = \{x_0,x_2\}$, $U_1=\{x_1,x_2\}$, and
$U_2=\{x_3,x_4\}$. With $\mathcal{U}(X) = \{U_0,U_1,U_2\}$, we
find $\hat\xi: \mathcal{U}(X)\times \Sigma_o \rightarrow \mathcal{U}(X)$,
and then the partial-observation supervisor $\bf SUPO$ is constructed.
Note that in $\bf SUPO$, only observable events lead to state changes.
Notation: a circle with $\rightarrow$ denotes the
initial state, and a double circle denotes a marker state; this
notation will be used throughout.} \label{fig:Example:GSUP}
\end{figure}

The transition function associated with
$\mathcal{U}(X)$ is $\hat\xi : \mathcal{U}(X) \times \Sigma_o
\rightarrow \mathcal{U}(X)$ given by
\begin{align} \label{eq:Utransit}
\hat\xi(U,\sigma) = \bigcup \{\xi(x,u_1 \sigma u_2) | x\in U,
u_1,u_2\in \Sigma_{uo}^*\}.\footnotemark
\end{align}
\footnotetext{Let $U = U(s)$ for some string $s \in L({\bf SUP})$;
then by definition of uncertainty set, $\bigcup \{\xi(x,u_1 \sigma
u_2) | x\in U, u_1,u_2\in \Sigma_{uo}^*\} = \{\xi(x_0,s'u_1 \sigma
u_2) | \xi(x_0,s')!, Ps' = Ps, u_1,u_2\in \Sigma_{uo}^*\} =
\{\xi(x_0,s'') | \xi(x_0,s'')!, Ps'' = P(s\sigma)\} = U(s\sigma) \in
\mathcal{U}(X)$.} If there exist $u_1,u_2\in \Sigma_{uo}^*$ such
that $\xi(x,u_1\sigma u_2)!$, then $\hat\xi(U,\sigma)$ is defined,
denoted as $\hat\xi(U,\sigma)!$.
With $\mathcal{U}(X)$ and $\hat\xi$, define the
\emph{partial-observation monolithic supervisor}
\cite{Wonham:2015a,CassandrasLafortune:08}
\begin{align}\label{eq:SUPO}
{\bf SUPO} = (\mathcal{U}(X),\Sigma_o,\hat{\xi},U_0,\mathcal{U}_m)
\end{align}
where $U_0=U(\epsilon)$ and $\mathcal{U}_m = \{U \in \mathcal{U}(X)
| U \cap X_m \neq \emptyset\}$. It is known
\cite{Wonham:2015a,CassandrasLafortune:08} that $L({\bf SUPO}) =
P(L({\bf SUP}))$ and $L_m({\bf SUPO}) = P(L_m({\bf SUP}))$.\footnote{
To ensure that the partial-observation supervisor $\bf SUPO$
is feasible, it is necessary to selfloop each state $U$ of $\hat\xi$ by
$\sigma = \Sigma_{uo}$ exactly when there exists $x \in U$ such
that $\xi(x,\sigma)!$. }
For an example of uncertainty set and partial-observation monolithic
supervisor, see Fig.~\ref{fig:Example:GSUP}.

Now let $U \in \mathcal{U}(X)$, $x \in U$ be any state in ${\bf SUP}$ and $\alpha \in
\Sigma_c$ be a controllable event. We say that (1) $\alpha$ is {\it
enabled} at $x \in U$ if \[\xi(x,\alpha)!;\] (2) $\alpha$ is {\it
disabled} at $x \in U$ if
\[\neg \xi(x,\alpha)!  ~\text{and}(\exists s \in \Sigma^*)\xi(x_0,s)=x ~\&~ \hat\xi(U_0,Ps) = U ~\&~
\delta(q_0,s\alpha)!;\] (3) $\alpha$ is {\it not defined} at $x\in U$
if \[\neg \xi(x,\alpha)! ~\text{and} (\forall s \in \Sigma^*)\xi(x_0,s)=x ~\&~\hat\xi(U_0,Ps) = U \Rightarrow
\neg \delta(q_0,s\alpha)!.\]

Under partial observation, the control actions 
after string $s \in L({\bf SUP})$ do not depend on the individual state
$\xi(x_0,s) \in X$, but just on the uncertainty set $U(s) \in
\mathcal{U}(X)$ (i.e. the state of $\bf SUPO$). Since the
language $L_m({\bf SUP})$ is (relatively) observable, the following is true.

\begin{lemma} \label{lem:Tproperty}
Given ${\bf SUP}$ in (\ref{eq:monosup}), let
$U \in \mathcal{U}(X)$, $x \in U$, and $\alpha \in \Sigma_c$. If
$\alpha$ is enabled at $x$, then
for all $x' \in U$, either $\alpha$ is also enabled at $x'$, or
$\alpha$ is not defined at $x'$.
On the other hand, if $\alpha$ is disabled at $x$,
then for all $x' \in U$, either $\alpha$ is also disabled at $x'$,
or $\alpha$ is not defined at $x'$. 
\end{lemma}

{\it Proof.}
By $x \in U \in \mathcal{U}(X)$, there exists $s\in L({\bf SUP})$ such that $\xi(x_0,s) = x$ and $U(s) = U$.
Suppose that  $\sigma \in \Sigma_c$ is enabled at $x \in U$, i.e. $\xi(x,\sigma)!$; it follows that $\xi(x,s\sigma)!$, i.e. $s\sigma \in L({\bf SUP})$. Now let $x' \in U=U(s)$. According to the subset construction algorithm, there must exist
$s' \in L({\bf SUP})$ such that $\xi(x_0,s') = x'$ (i.e. $s' \in L({\bf SUP})$) and $Ps' = Ps$.
At state $x'$, either (i) $\xi(x',\sigma)!$, or (ii) $\neg \xi(x',\sigma)!$.
Case (i) means that $\sigma$ is enabled at $x' \in U$. In case (ii), we claim that
$s'\sigma \notin L({\bf G})$, i.e. $\sigma$ is not defined at $x' \in U$.
To see this, assume on the contrary that $s'\sigma \in L({\bf G})$. Then we have $Ps'=Ps$, $s' \in L({\bf SUP})$, $s'\sigma \in L({\bf G})$, $s' \sigma \notin L({\bf SUP})$, and $s\sigma \in L({\bf SUP})$. This implies that $L_m({\bf SUP})$ is not observable, which is a contradiction to the definition of $L_m({\bf SUP})$ in (5).
Therefore, in case (ii), $\sigma$ is not defined at $x' \in U$ after all.

The second statement can be proved by a similar argument.

\hfill $\square$


\subsection{Localization Procedure}

The procedure of partial-observation localization proceeds similarly
to \cite{CaiWonham:2010a}, but is based on the set $\mathcal{U}(X)$
of the uncertainty sets and its associated transition function $\hat
\xi$, i.e. based on the partial-observation monolithic supervisor
{\bf SUPO} in (\ref{eq:SUPO}).

First, consider the following four functions which capture the
control and marking information on the uncertainty sets. Fix a
controllable event $\alpha \in \Sigma_c$. Define
$E_\alpha:\mathcal{U}(X) \rightarrow \{0,1\}$ according to
\begin{equation*} \label{eq:ET1}
\begin{split}
(\forall U \in \mathcal{U}(X))~
E_\alpha(U) &= \left\{
  \begin{array}{ll}
    1, & \hbox{if $(\exists x \in U) \xi(x,\alpha)!$,} \\
    0, & \hbox{otherwise.}
  \end{array}
\right.\\
\end{split}
\end{equation*}
Thus $E_\alpha(U) = 1$ if event $\alpha$ is enabled at
some state $x \in U$. Then by Lemma~\ref{lem:Tproperty} at any other
state $x'\in U$, $\alpha$ is either enabled or not defined. Also
define $D_\alpha:\mathcal{U}(X) \rightarrow \{0,1\}$ according to
\begin{equation*} \label{eq:ET2}
\begin{split}
&(\forall U \in \mathcal{U}(X)) \\
&D_\alpha(U)= \left\{
  \begin{array}{ll}
    1, & \hbox{if $(\exists x \in U) \neg\xi(x,\alpha)! \ \& (\exists s\in \Sigma^*)$}\\
     & \hbox{$\xi(x_0,s)= x ~\&~ \hat\xi(U_0,Ps) = U ~\&~\delta(q_0,s\alpha)!$,} \\
    0, & \hbox{otherwise.}
  \end{array}
\right.\\
\end{split}
\end{equation*}
Hence $D_\alpha(U) = 1$ if $\alpha$ is disabled at some
state $x \in U$. Again by Lemma~\ref{lem:Tproperty} at any other
state $x'\in U$, $\alpha$ is either disabled or not defined.

Consider the example displayed in
Fig.~\ref{fig:Example:GSUP}. The control actions include (i) enabling events 1, 3 at state $x_0$, event 5 at
state $x_3$; and (ii) disabling event 3 at state $x_1$, event 5 at state $x_2$.
For the uncertainty set $U_0 = \{x_0, x_2\}$, $E_3(U_0) = 1$ because
event 3 is enabled at state $x_0 \in U_0$; note that event 3 is not
defined at the other state $x_2 \in U_0$. For the uncertainty set
$U_1 =\{x_1,x_2\}$, $D_3(U_1)=1$ because event $3$ is disabled at state
$x_1 \in U_1$; also note that event 3 is not defined at state $x_2 \in
U_1$.

Next, define $M: \mathcal{U}(X) \rightarrow \{0,1\}$ according to
\begin{equation*} \label{eq:M}
\begin{split}
(\forall U \in \mathcal{U}(X)) ~ M(U) = \left\{
  \begin{array}{ll}
    1, & \hbox{if $U \in \mathcal {U}_m$,} \\
    0, & \hbox{otherwise.}
  \end{array}
\right.\\
\end{split}
\end{equation*}
Thus $M(U)=1$ if $U$ is marked in {\bf SUPO} (i.e. $U$ contains a
marker state of $\bf SUP$). Finally define $T:\mathcal{U}(X)
\rightarrow \{0,1\}$ according to
\begin{equation*} \label{eq:T}
\begin{split}
(\forall & U \in\mathcal{U}(X))\\
&T(U)=\left\{
  \begin{array}{ll}
    1, & \hbox{if $(\exists s \in \Sigma^*)\xi(x_0,s) \in U ~\&~$} \\
      & \hbox{$ \hat\xi(U_0,Ps) = U~\&~\delta(q_0,s)\in Q_m$,} \\
    0, & \hbox{otherwise.}
  \end{array}
\right.\\
\end{split}
\end{equation*}
So $T(U) = 1$ if $U$ contains some state that corresponds (via a
string $s$) to a marker state of $\bf G$.

With the above four functions capturing control and marking
information of the uncertainty sets in $\mathcal{U}(X)$, we define
the \emph{control consistency relation} $\mathcal{R}_\alpha
\subseteq \mathcal{U}(X) \times \mathcal{U}(X)$ as follows.
\begin{definition} \label{def:controlconsist}
For $U,U' \in \mathcal{U}(X)$, we say that $U$ and $U'$ are {\it
control consistent} with respect to $\alpha$,
written $(U,U')\in \mathcal{R}_\alpha$, if
\begin{align*}
\mbox{(i)}~~ & E_\alpha(U)\cdot D_\alpha(U') = 0 = E_\alpha(U')\cdot D_\alpha(U),\\
\mbox{(ii)}~~ &T(U)=T(U') \Rightarrow M(U)=M(U').
\end{align*}
\end{definition}
Thus a pair of uncertainty sets $(U,U')$ satisfies $(U,U')\in
\mathcal{R}_\alpha$ if (i) event $\alpha$ is enabled at at least one
state of $U$, but is not disabled at any state of $U'$, and vice versa;
(ii) $U$, $U'$ both contain marker states of $\bf SUP$ (resp. both
do not contain) provided that they both contain states corresponding
to some marker states of $\bf G$ (resp. both do not contain).

For example, in Fig.~\ref{fig:Example:GSUP}, for event
$3$ we have:
\[\begin{array}{ccccc}
   & E_3 & D_3 & M & T \\
  U_0 & 1 & 0 & 1 & 1 \\
  U_1 & 0 & 1 & 0 & 0 \\
  U_2 & 0 & 0 & 0 & 0
\end{array}\]
Hence $(U_0, U_2) \in \mathcal{R}_3$, $(U_2, U_1) \in
\mathcal{R}_3$, and $(U_0, U_1) \notin \mathcal{R}_3$. From this
example we see that $\mathcal{R}_\alpha$ is generally not
transitive, and thus not an equivalence relation. This fact leads to
the following definition of a {\it partial-observation control
cover}.


\begin{definition} \label{def:controlcover}
Let $I$ be some index set, and $\mathcal {C}_\alpha =
\{\mathcal{U}_i \subseteq\mathcal{U}(X) | i\in I\}$ be a cover on
$\mathcal{U}(X)$. We say that $\mathcal {C}_\alpha$ is a {\it
partial-observation control cover} with respect to $\alpha$ if
\begin{align*}
\mbox{(i)}~~ & (\forall i \in I, \forall U,U' \in \mathcal{U}_i)~ (U,U') \in \mathcal{R}_\alpha,\\
\mbox{(ii)}~~ & (\forall i \in I, \forall \sigma \in \Sigma_o) (\exists U\in
\mathcal{U}_i) ~\hat\xi(U,\sigma)! \Rightarrow
 \big[(\exists j \in I)\\
 &~~~~~~~~~~~~~~~(\forall U'\in \mathcal{U}_i)~\hat\xi(U',\sigma)!
                  \Rightarrow \hat\xi(U',\sigma) \in \mathcal{U}_j\big].
\end{align*}
\end{definition}
A partial-observation control cover $\mathcal {C}_\alpha$ lumps the
uncertainty sets $U \in \mathcal{U}(X)$ into (possibly overlapping)
{\it cells} $\mathcal{U}_i \in \mathcal {C}_\alpha$, $i\in I$,
according to (i) the uncertainty
sets $U$ that reside in the same cell $\mathcal{U}_i$ must be
pairwise control consistent, and (ii) for every
observable event $\sigma\in \Sigma_o$, the uncertainty set that
is reached from any uncertainty set $U'\in \mathcal{U}_i$ by a
one-step transition $\sigma$ must be covered by the same cell
$\mathcal{U}_j$. Inductively, two uncertainty sets $U$ and $U'$
belong to a common cell of $\mathcal{C}_\alpha$ if and only if $U$
and $U'$ are control consistent, and two future
uncertainty sets that can be reached respectively from $U$ and $U'$
by a given observable string are again control
consistent.

The partial-observation control cover $\mathcal {C}_\alpha$ differs
from its counterpart in \cite{CaiWonham:2010a} in two
aspects. First, $\mathcal {C}_\alpha$ is defined on
$\mathcal{U}(X)$, not on $X$; this is due to state uncertainty
caused by partial observation. Second, in condition (ii) of
$\mathcal {C}_\alpha$ only observable events in $\Sigma_o$ are
considered, not $\Sigma$; this is to generate partial-observation
local controllers whose state transitions are triggered only by
observable events. We call $\mathcal{C}_\alpha$ a {\it
partial-observation control congruence} if $\mathcal{C}_\alpha$
happens to be a partition on $\mathcal{U}(X)$, namely its cells are
pairwise disjoint.

Having defined a partial-observation control cover
$\mathcal{C}_\alpha$ on $\mathcal{U}(X)$, we construct a generator
${\bf J}_\alpha = (I,\Sigma_o,\zeta_\alpha,i_0,I_m)$ defined over $\Sigma_o$
and a control function
$\psi_\alpha:I\rightarrow \{0,1\}$ as follows:
\begin{align}
\mbox{(i)}~~ & i_0 \in I ~\text{such that}~  (\exists U \in \mathcal{U}_{i_0})x_0 \in U;\label{eq:sub1:construct}\\
\mbox{(ii)}~~ & I_m := \{i \in I |  (\exists U \in \mathcal{U}_i) X_m\cap U \neq \emptyset \};\label{eq:sub2:construct}\\
\mbox{(iii)}~~ & \zeta_\alpha:I\times\Sigma_o\rightarrow I ~\text{with}~
\zeta_\alpha(i,\sigma)=j \notag\\
       &\text{if}~ (\exists U \in \mathcal{U}_i)~\hat{\xi}(U,\sigma)\in \mathcal{U}_j;
\label{eq:sub3:construct}\\
\mbox{(iv)}~~ &\psi_\alpha(i) = 1 ~\text{iff}~ (\exists U \in
\mathcal{U}_i)~ E_\alpha(U) = 1. \label{eq:sub4:construct}
\end{align}
The control function $\psi_\alpha(i) = 1$ means that event $\alpha$
is enabled at state $i$ of ${\bf J}_\alpha$. Note that owing to
cell overlapping, the choices of $i_0$ and $\zeta_\alpha$ may not be
unique, and consequently ${\bf J}_\alpha$ may not be unique. In that
case we pick an arbitrary instance of ${\bf J}_\alpha$.

Finally we define the {\it partial-observation local controller} ${\bf
LOC}_\alpha = (Y_\alpha,\Sigma_\alpha,\eta_\alpha,y_{0,\alpha},Y_{m,\alpha})$
as follows.

\noindent (i) $Y_\alpha = I$, $y_{0,\alpha} = i_0$, and
$Y_{m,\alpha} = I_m$.
Thus the control function $\psi_\alpha$ is $\psi_\alpha: Y_\alpha\rightarrow\{0,1\}$.

\noindent (ii) $\Sigma_\alpha = \{\alpha\} \cup
\Sigma_{com,\alpha}$, where
\begin{align} \label{eq:comevent}
\Sigma_{com,\alpha} := \{\sigma \in \Sigma_o \setminus \{\alpha\} \
| \ (\exists i,j \in I)  i \neq j,\ \zeta_\alpha(i,\sigma)=j\}
\end{align}
Thus $\Sigma_{com,\alpha}$ is the set of observable events that are
not merely selfloops in ${\bf J}_\alpha$. It holds by definition
that $\{\alpha\} \subseteq \Sigma_\alpha \subseteq \Sigma_o \cup
\{\alpha\}$, and $\Sigma_{com,\alpha}$ contains the events of other
local controllers that need to be communicated to ${\bf
LOC}_\alpha$.

\noindent (iii) If $\alpha \in \Sigma_o$, then $\eta_\alpha :=
\zeta_\alpha|_{Y_\alpha \times \Sigma_\alpha} : Y_\alpha \times
\Sigma_\alpha \rightarrow Y_\alpha$, i.e. $\eta_\alpha$ is the restriction
of $\zeta_\alpha$ to $Y_\alpha \times \Sigma_\alpha$. If $\alpha \in
\Sigma_{uo}$, first obtain $\eta_\alpha := \zeta_\alpha|_{Y_\alpha \times
\Sigma_\alpha}$ and then add $\alpha$-selfloops $\eta_\alpha(y,\alpha)=y$
to those $y\in Y_\alpha$ with $\psi_\alpha(y) = 1$.

\begin{lemma} \label{lem:loc}
The generator ${\bf LOC}_\alpha$ is a partial-observation local
controller for $\alpha$, i.e. (\ref{eq:loc1}) and (\ref{eq:loc2})
hold.
\end{lemma}

We postpone the proof of Lemma~\ref{lem:loc} after our main result,
Theorem~\ref{thm:equ}, in the next subsection.

\begin{figure}[!t]
\centering
    \includegraphics[scale = 0.4]{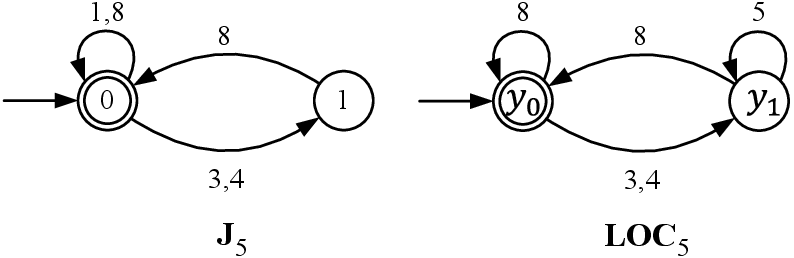}
\caption{Generator ${\bf J}_5$ and partial-observation local
controller ${\bf LOC}_5$. In ${\bf J}_5$, state 0 corresponds to
cell $\{U_0,U_1\}$ of the control cover $\mathcal{C}_5 = \{\{U_0, U_1\},\{U_2\}\}$
while state 1 corresponds to cell $\{U_2\}$. From
${\bf J}_5$ to ${\bf LOC}_5$, (i) the 5-selfloop at state $y_1$ is
added because $\psi_5(\{U_2\}) = 1$ and event 5 is unobservable, and
(ii) event 1 is removed since it is merely a selfloop in ${\bf
J}_5$ and thus its occurrences will not affect the enablement/disablement
of event 5.} \label{fig:Example:Gen5}
\end{figure}

Consider again the example displayed in Fig.~\ref{fig:Example:GSUP}.
We construct a partial-observation local controller ${\bf LOC}_5$
for the unobservable controllable event $5$. For event $5$ we have:
\[\begin{array}{ccccc}
   & E_5 & D_5 & M & T \\
  U_0 & 0 & 1 & 1 & 1 \\
  U_1 & 0 & 1 & 0 & 0 \\
  U_2 & 1 & 0 & 0 & 0
\end{array}\]
Hence $(U_0,U_1) \in \mathcal{R}_5$, $(U_0,U_2) \notin
\mathcal{R}_5$, and $(U_1,U_2) \notin \mathcal{R}_5$. Further,
because $\hat\xi(U_0,8)=\hat\xi(U_1,8)=U_0$, $U_0$ and $U_1$ can be
put in the same cell; but $U_0$ and $U_2$ cannot, nor can $U_1$ and
$U_2$. So we get a partial-observation control cover $\mathcal {C}_5
= \{\{U_0,U_1\},\{U_2\}\}$. From this control cover, we construct a
generator ${\bf J}_5$ as shown in Fig.~\ref{fig:Example:Gen5}, and a
control function $\psi_5$ such that $\psi_5(\{U_0,U_1\})=0$ and
$\psi_5(\{U_2\}) = 1$ because $E_5(U_2)=1$. Finally the
partial-observation local controller ${\bf LOC}_5$ is constructed
from the generator ${\bf J}_5$ by adding the 5-selfloop at state
$y_1$ because $\psi_5(\{U_2\}) = 1$ and $5$ is unobservable, and
removing event 1 since it is merely a selfloop in ${\bf J}_5$ (see
Fig.~\ref{fig:Example:Gen5}).

\subsection{Main Result}

By the same procedure as above, we construct a set of
partial-observation local controllers ${\bf LOC}_\alpha$, one for
each controllable event $\alpha \in \Sigma_c$. We shall verify that
these local controllers collectively achieve the same controlled
behavior as represented by ${\bf SUP}$ in (\ref{eq:monosup}).

\begin{theorem} \label{thm:equ}
The set of partial-observation local controllers $\{{\bf
LOC}_\alpha|\alpha \in \Sigma_c\}$ is a solution to the
Partial-Observation Supervisor Localization Problem, i.e.
\begin{align}
   L({\bf G}) \cap L({\bf LOC})  &= L({\bf SUP}) \label{eq:sub1:equv}\\
   L_m({\bf G})\cap L_m({\bf LOC}) &= L_m({\bf SUP}) \label{eq:sub2:equv}
\end{align}
where $L({\bf LOC}) = \mathop \bigcap\limits_{\alpha \in \Sigma_{c}}
   P_\alpha^{-1}L({\bf LOC}_{\alpha})$ and $L_m({\bf LOC}) = \mathop \bigcap\limits_{\alpha \in \Sigma_{c}}
   P_\alpha^{-1}L_m({\bf LOC}_{\alpha})$.
\end{theorem}

\noindent {\it Proof}: 
%
%
First we show that $L_m({\bf SUP}) \subseteq L_m({\bf G}) \cap L_m({\bf LOC})$.
It suffices to show $(\forall \alpha \in \Sigma_c)~L_m({\bf SUP})\subseteq
P_\alpha^{-1}L_m({\bf LOC}_\alpha)$. Let $\alpha \in \Sigma_c$ and $s \in L_m({\bf SUP})$;
we must show $s \in P_\alpha^{-1}L_m({\bf LOC}_\alpha)$.
Write $Ps = \sigma_0,...,\sigma_n$; then $Ps \in PL_m({\bf SUP})$. According to
the definition of uncertainty set, there exist $U_0, ..., U_n \in \mathcal{U}(X)$
such that
\[\hat\xi(U_j,\sigma_j) = U_{j+1}, j = 0, ..., n-1.\]
Then by the definition of $\mathcal{C}_\alpha$ and $\zeta_\alpha$,
for each $j = 0, ..., n - 1$, there exist
$i_j, i_{j+1} \in I$ such that
\[U_j \in \mathcal{U}_{i_j} ~\&~ U_{j+1} \in \mathcal{U}_{i_{j+1}} ~\&~ \zeta_\alpha(i_j,\sigma_j)=i_{j+1}.\]
So $\zeta_\alpha(i_0,\sigma_0...\sigma_n)!$, i.e. $\zeta_\alpha(i_0, Ps)!$. Let $i_n = \zeta_\alpha(i_0,Ps)$;
then $U(Ps) \in \mathcal{U}_{i_n}$, and thus $\xi(x_0,s) \in U(Ps) \cap X_m$.
So $i_n \in I_m$, i.e. $Ps \in L_m({\bf J}_\alpha)$. Let $P_\alpha':\Sigma_o^*\rightarrow\Sigma_\alpha^*$
be the natural projection as defined in (\ref{eq:natpro}); then
$P_\alpha(s) = P_\alpha'(Ps) \in P_\alpha'L_m({\bf J}_\alpha) = L_m({\bf LOC}_\alpha)$.
Hence $s \in P_\alpha^{-1}L_m({\bf LOC}_\alpha)$.

Now that we have shown
$L_m({\bf SUP}) \subseteq L_m({\bf G})\cap L_m({\bf LOC})$, it follows that
\begin{align*}
L({\bf SUP}) & = \overline{L_m({\bf SUP})}\\
             & \subseteq \overline{L_m({\bf G})\cap L_m({\bf LOC})}\\
             & \subseteq \overline{L_m({\bf G})} \cap \overline{L_m({\bf LOC})}\\
             & \subseteq L({\bf G}) \cap \mathop \bigcap\limits_{\alpha \in \Sigma_{c}}
                    P_\alpha^{-1}\overline{L_m({\bf LOC}_{\alpha})}\\
             & \subseteq L({\bf G}) \cap \mathop \bigcap\limits_{\alpha \in \Sigma_{c}}
                    P_\alpha^{-1}L({\bf LOC}_{\alpha}) \\
             & = L({\bf G}) \cap L({\bf LOC})
\end{align*}
so $L({\bf SUP}) \subseteq L({\bf G}) \cap L({\bf LOC})$.

Next, we prove  $L({\bf G}) \cap L({\bf LOC}) \subseteq L({\bf SUP})$, by induction
on the length of strings.

For the base case, as it was assumed that $L_m({\bf SUP})$ is
nonempty, it follows that the languages $L({\bf G})$, $L({\bf LOC})$
and $L({\bf SUP})$ are all nonempty, and as they are closed, the
empty string $\epsilon$ belongs to each.

For the inductive step, suppose that $s \in L({\bf G}) \cap L({\bf
LOC})$ implies $s \in L({\bf SUP})$, and $s\alpha \in L({\bf G})
\cap L({\bf LOC})$ for an arbitrary event $\alpha \in \Sigma$; we
must show that $s\alpha \in L({\bf SUP})$. If $\alpha \in \Sigma_u$,
then $s\alpha \in L({\bf SUP})$ because $L_m({\bf SUP})$ is
controllable. Otherwise, we have $\alpha \in \Sigma_c$ and there
exists a partial-observation local controller ${\bf LOC}_\alpha$ for
$\alpha$. It follows from $s\alpha \in L({\bf LOC})$ that $s\alpha
\in P_\alpha^{-1}L({\bf LOC}_\alpha)$ and $s \in P_\alpha^{-1}L({\bf
LOC}_\alpha)$. So $P_\alpha(s\alpha) \in L({\bf LOC}_\alpha)$ and
$P_\alpha (s) \in L({\bf LOC}_\alpha)$, namely,
$\eta_\alpha(y_0,P_\alpha(s\alpha))!$ and $\eta_\alpha(y_0,P_\alpha(s))!$. Let $y
:= \eta_\alpha(y_0,P_\alpha(s))$; then $\eta_\alpha(y,\alpha)!$ (because
$\alpha \in \Sigma_\alpha$). Since $\alpha$
may be observable or unobservable, we consider the following two
cases.

Case (1) $\alpha \in \Sigma_{uo}$. It follows from the construction
(iii) of ${\bf LOC}_\alpha$ that $\eta_\alpha(y,\alpha)!$ implies that for
the state $i \in I$ of the generator ${\bf J}_\alpha$ corresponding
to $y$ (i.e. $i = \zeta_\alpha(i_0,P(s))$), there holds $\psi_\alpha(i) =
1$. By the definition of $\psi_\alpha$ in (\ref{eq:sub4:construct}),
there exists an uncertainty set $U \in \mathcal{U}_i$ such that
$E_\alpha(U) = 1$. Let $U' = \hat\xi(U_0,Ps)$; by (\ref{eq:sub3:construct})
and $i = \zeta_\alpha(i_0, Ps)$, $U'\in \mathcal{U}_i$. According to (\ref{eq:Utransit}),
$\xi(x_0,s) \in U'$.  Since $U$ and $U'$ belong to the same cell
$\mathcal{U}_i$, by the definition of partial-observation control
cover they must be control consistent, i.e. $(U,U')\in
\mathcal{R}_\alpha$. Thus $E_\alpha(U)\cdot D_\alpha(U')=0$, which
implies $D_\alpha(U') = 0$. The latter means that for all states $x
\in U'$, either (i) $\xi(x,\alpha)!$ or (ii) for all $t \in
\Sigma^*$ with $\xi(x_0,t) = x$, $\delta(q_0,t\alpha)$ is not
defined. Note that (ii) is impossible for $\xi(x_0,s) \in U'$,
because $s\alpha \in L({\bf G})$. Thus by (i),
$\xi(\xi(x_0,s),\alpha)!$, and therefore $s\alpha \in L({\bf SUP})$.

Case (2) $\alpha \in \Sigma_o$. In this case, for the state $i \in
I$ of the generator ${\bf J}_\alpha$ corresponding to $y$ (i.e. $i =
\zeta_\alpha(i_0,P(s))$), there holds $\zeta_\alpha(i,\alpha)!$. By the definition
of $\zeta_\alpha$ in (\ref{eq:sub3:construct}), there exists an uncertainty
set $U \in \mathcal{U}_i$ such that $\hat\xi(U,\alpha)!$, i.e.
$E_\alpha(U) = 1$. The rest of the proof is identical to Case (1) above,
and we conclude that $s\alpha \in L({\bf SUP})$ in this case as well.

Finally we show $L_m({\bf G}) \cap L_m({\bf LOC}) \subseteq L_m({\bf SUP})$.
Let $s \in L_m({\bf G}) \cap L_m({\bf LOC})$; we must show that $s \in L_m({\bf SUP})$.
Since $L_m({\bf G})\cap L_m({\bf LOC}) \subseteq L({\bf G})\cap L({\bf LOC}) \subseteq L({\bf SUP})$,
we have $s \in L({\bf SUP})$, which implies that for all $\alpha \in \Sigma_c$,
$i_n = \zeta_\alpha(i_0, Ps)$ and $U(Ps) \in \mathcal{U}_{i_n}$.
In addition, $s \in L_m({\bf LOC})$ implies that $s \in P_\alpha^{-1} L_m({\bf LOC}_\alpha)$;
so $P_\alpha s \in L_m({\bf LOC}_\alpha)$, i.e. $\eta_\alpha(y_0, P_\alpha s) \in Y_m$.
Since $P_\alpha s = P_\alpha'(Ps)$, $\eta_\alpha(y_0, P_\alpha s)$ corresponds to $\zeta_\alpha(i_0, Ps) = i_n$;
so $i_n \in I_m$ (because $Y_m = I_m$). Therefore, there exists $U' \in \mathcal{U}_{i_n}$
such that $X_m \cap U' \neq \emptyset$. Then $M(U') = 1$ and thus $T(U') = 1$.
By $s \in L_m({\bf G})$, we have $T(U(Ps)) = 1$.
Now we have that both $U'$ and $U(Ps)$ are in $\mathcal{U}_{i_n}$, i.e.
$(U', U(Ps)) \in R_\alpha$. Consequently $M(U(Ps)) = M(U') = 1$.
Hence $U(Ps) \cap X_m \neq \emptyset$. Let $x = U(Ps) \cap X_m$;
then there must exist $t \in L_m({\bf SUP})$ such that $x = \xi(x_0,t)$ and $Pt = Ps$.
Now, since $L_m({\bf SUP})$ is observable, by $s \in L_m({\bf G})$ we have $s \in L_m({\bf SUP})$.


\hfill $\square$

\begin{figure}[!t]
\centering
    \includegraphics[scale = 0.16]{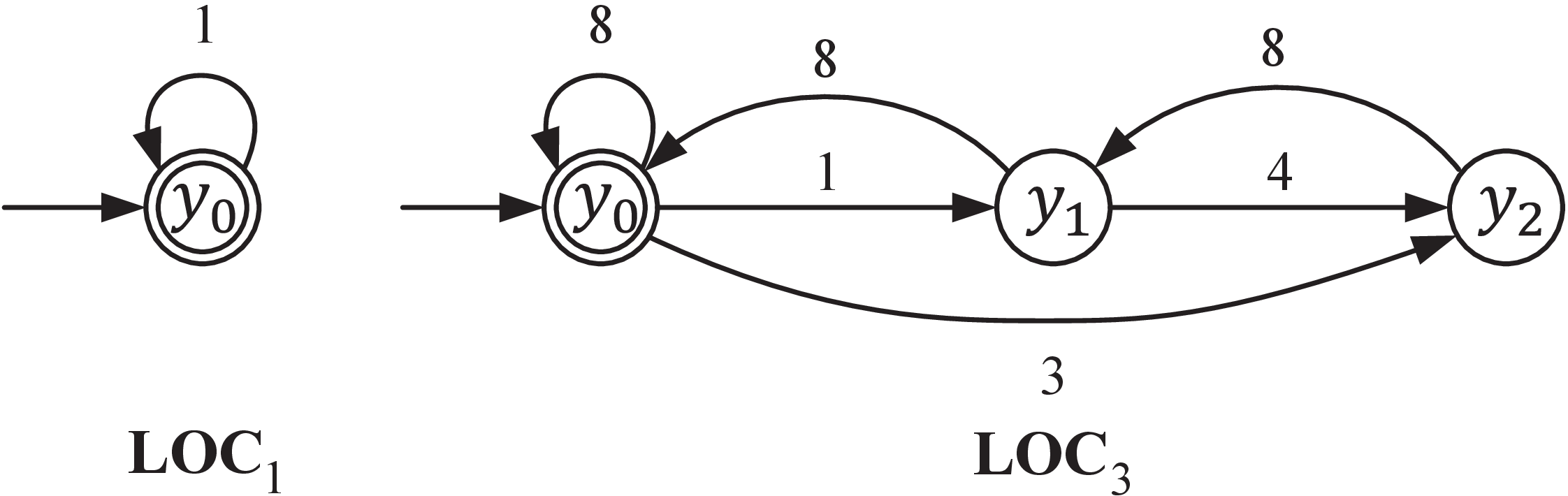}
\caption{Partial-observation local controllers ${\bf LOC}_1$ and
${\bf LOC}_3$. For ${\bf LOC}_1$, since event 1 is not disabled at
any uncertainty set, the partial-observation control cover $\mathcal
{C}_1 = \{\{U_0,U_1,U_2\}\}$ and ${\bf LOC}_1$ has just one state at
which event 1 is enabled. For ${\bf LOC}_3$, note that the
uncertainty sets $U_0$ and $U_2$ (corresponding to states $y_0$ and
$y_2$ respectively) are not in a common cell of $\mathcal {C}_3$
even though $U_0$ and $U_2$ are control consistent; this is because
after an $8$-transition, they will arrive in $U_0$ and $U_1$
respectively, but $U_0$ and $U_1$ are not control consistent. Thus
$\mathcal {C}_3 = \{\{U_0\},\{U_1\},\{U_2\}\}$ and ${\bf LOC}_3$ has
three states.} \label{fig:Example:LOCs}
\end{figure}

For the example in Fig.~\ref{fig:Example:GSUP}, we construct
partial-observation local controllers ${\bf LOC}_1$ and ${\bf
LOC}_3$ for the (observable) controllable events 1 and 3
respectively, as displayed in Fig.~\ref{fig:Example:LOCs}. It is
then verified that the collective controlled behavior of these local
controllers (${\bf LOC}_1$, ${\bf LOC}_3$, and ${\bf LOC}_5$) is
identical to ${\bf SUP}$ (in the sense of (\ref{eq:sub1:equv}) and
(\ref{eq:sub2:equv})). \footnote{This can be verified by TCT procedures as follows.
First, compute ${\bf TEST} = Sync({\bf G},{\bf LOC1}, {\bf LOC3}, {\bf LOC5})$,
i.e. ${\bf TEST} = {\bf G} || {\bf LOC1}||{\bf LOC3}||{\bf LOC5}$.
Then it is verified by $true = Isomorph({\bf TEST}, {\bf SUP})$ that
$L_m({\bf TEST}) = L_m({\bf SUP})$ and $L({\bf TEST}) = L({\bf SUP})$.}

Finally, we provide the proof of Lemma~\ref{lem:loc}.

{\it Proof of Lemma~\ref{lem:loc}.} We must prove (\ref{eq:loc1}) and
(\ref{eq:loc2}).

First, for ($\Rightarrow$) of Eq. (\ref{eq:loc1}), let $P_\alpha(s)\alpha \in L({\bf LOC}_\alpha)$,
$s\alpha \in L({\bf G})$ and $s \in L({\bf SUP})$; we must prove
that $s\alpha \in L({\bf SUP})$. It is derived from $P_\alpha(s)\alpha \in L({\bf LOC}_\alpha)$,
that $P_\alpha(s) \in L({\bf LOC}_\alpha)$, because $L({\bf LOC}_\alpha)$ is prefix-closed.
Let $y := \eta_\alpha(y_{0,\alpha},P_\alpha(s))!$; by $P_\alpha(s)\alpha \in L({\bf LOC}_\alpha)$,
$\eta_\alpha(y,\alpha)!$. The rest of the proof is identical
to the inductive case of proving ($\subseteq$) of (\ref{eq:sub1:equv}),
and we conclude that $s\alpha \in L({\bf SUP})$.




Next, for ($\Leftarrow$) of Eq. (\ref{eq:loc1}), let $s\alpha \in L({\bf SUP})$;
$s \in L({\bf SUP})$ and $s\alpha \in L({\bf G})$ are immediate, and it is left to show
that $P_\alpha(s)\alpha \in L({\bf LOC}_\alpha)$. By $s\alpha \in L({\bf SUP})$
and (\ref{eq:sub1:equv}), we have for all $\sigma \in \Sigma_c$,
$s\alpha \in P^{-1}_\sigma L({\bf LOC}_\sigma)$. Because $\alpha \in \Sigma_c$,
we have $s\alpha \in P^{-1}_\alpha L({\bf LOC}_\alpha)$,
and thus $P_\alpha(s\alpha) \in L({\bf LOC}_\alpha)$.
According to the definition of $\Sigma_\alpha$, $\{\alpha\} \subseteq \Sigma_\alpha$.
Hence, $P_\alpha(s)\alpha = P_\alpha(s\alpha) \in L({\bf LOC}_\alpha)$.


Finally, to prove (\ref{eq:loc2}), let $y, y' \in Y_\alpha$ and $\sigma \in \Sigma_o$
and assume that $y' = \eta_\alpha(y,\sigma)$ and $y \neq y'$; we prove that
$\sigma \in \Sigma_o$ by contradiction. Suppose that $\sigma \in \Sigma_{uo}$.
According to (\ref{eq:sub3:construct}), for all $i \in I$, $\zeta_\alpha(i,\sigma)$
is not defined. Further, according to the rule (iii)
of constructing ${\bf LOC}_\alpha$, (1) for all $y \in Y$, $\eta_\alpha(y,\sigma)$
is not defined, contradicting the assumption that $y' = \eta_\alpha(y,\alpha)$;
(2) the selfloop $\eta_\alpha(y,\alpha) = y$ is added to $\eta_\alpha$ when $\psi_\alpha(y) = 1$,
which, however, contradicts the assumption that $y \neq y'$.
So we conclude that $\sigma \in \Sigma_o$.

\hfill $\square$


\section{Partial-Observation Localization Algorithm and Transfer Line Example}

In this section, we adapt the supervisor localization
algorithm in \cite{CaiWonham:2010a} to compute the
partial-observation local controllers.

Let ${\bf SUP} = (X, \Sigma, \xi, x_0, X_m)$ be the controllable and
observable controlled behavior (as in (\ref{eq:monosup})), with
controllable $\Sigma_c$ and observable $\Sigma_o$. Fix $\alpha \in
\Sigma_c$. The algorithm in \cite{CaiWonham:2010a} would construct a
control cover on $X$. Here instead, owing to partial observation, we
first find the set $\mathcal {U}(X)$ of all uncertainty sets and
label it as
\begin{align*}
\mathcal {U}(X) = \{U_0,U_1,...,U_{n-1}\}.
\end{align*}
Also we calculate the transition function $\hat\xi:\mathcal{U}(X) \times
\Sigma^*_o \rightarrow \mathcal{U}(X)$. These steps are 
done by constructing the partial-observation monolithic supervisor
{\bf SUPO} as in (\ref{eq:SUPO})
\cite{Wonham:2015a,CassandrasLafortune:08}.

Next, we apply the localization algorithm in
\cite{CaiWonham:2010a} to construct a
partial-observation control cover $\mathcal {C}_\alpha$ on $\mathcal
{U}(X)$. Initially $\mathcal {C}_\alpha$ is set to be the singleton
partition on $\mathcal {U}(X)$, i.e.
\begin{align*}
\mathcal {C}_\alpha = \{\{U_0\},\{U_1\},...,\{U_{n-1}\}\}.
\end{align*}
Write $\mathcal {U}_i, \mathcal {U}_j$ for two cells in $\mathcal
{C}_\alpha$. Then the algorithm `merges' $\mathcal {U}_i, \mathcal
{U}_j$ into one cell if for every uncertainty set $U_i \in \mathcal
{U}_i$ and every $U_j \in \mathcal {U}_j$, $U_i$ and $U_j$, as well
as their corresponding future uncertainty sets reachable by
identical strings, are control consistent in
terms of $\mathcal {R}_\alpha$. The algorithm loops until all
uncertainty sets in $\mathcal {U}(X)$ are checked for control
consistency. We call this algorithm the {\it partial-observation localization
algorithm}.

Similar to \cite{CaiWonham:2010a}, the algorithm
terminates in a finite number of steps and results in a
partial-observation control {\it congruence} $\mathcal {C}_\alpha$ (i.e.
with pairwise disjoint cells). The complexity of the algorithm is
$O(n^4)$; since the size $n$ of $\mathcal {U}(X)$ is $n \leq
2^{|X|}$ in general, the algorithm is exponential in $|X|$.

\begin{figure}[!t]
\centering
    \includegraphics[scale = 0.4]{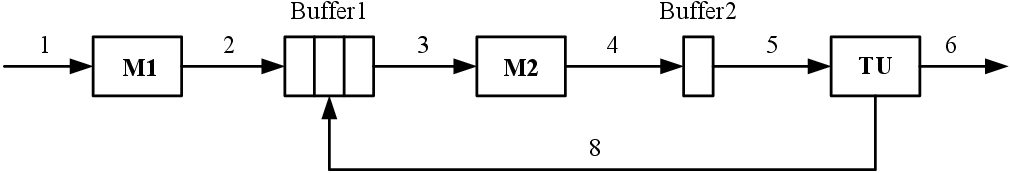}
\caption{Transfer Line: system configuration, with the set of
controllable events $\Sigma_c = \{1,3,5\}$} \label{fig:TL}
\end{figure}

In the following, we illustrate the above partial-observation
localization algorithm by a Transfer Line system $\bf TL$, as
displayed in Fig.~\ref{fig:TL}.  $\bf TL$ consists of two machines
$\bf M1$, $\bf M2$ followed by a test unit $\bf TU$; these agents
are linked by two buffers (Buffer1, Buffer2) with capacities of
three slots and one slot, respectively. We model the synchronous
product of $\bf M1$, $\bf M2$, and $\bf TU$ as
the plant to be controlled; the specification is to protect the two
buffers against overflow and underflow.

\begin{figure}[!t]
\centering
    \includegraphics[scale = 0.8]{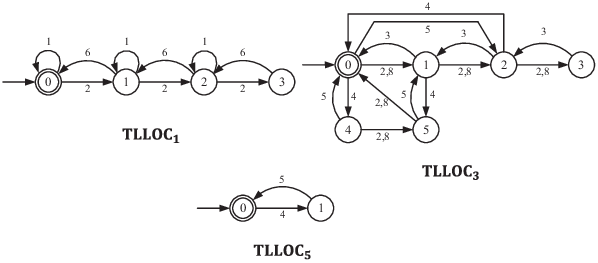}
\caption{Transfer Line: local controllers with full observation}
\label{fig:full_obs_LOC}
\end{figure}

For comparison purpose, we first present the local controllers
under full observation. By \cite{CaiWonham:2010a}, these controllers
are as displayed in Fig.~\ref{fig:full_obs_LOC}, and their control logic
is as follows.

${\bf TLLOC}_1$ for agent $\bf M1$ ensures that no more than three
workpieces can be processed in the material-feedback loop. This is realized by
counting the occurrences of event 2 (input a workpiece into the
loop) and event 6 (output a workpiece from the loop).

${\bf TLLOC}_3$ for agent $\bf M2$ guarantees no overflow or underflow
of the two buffers. This is realized by counting events 2, 8 (input a
workpiece into Buffer1), 3 (output a workpiece from Buffer1), 4
(input a workpiece into Buffer2), and 5 (output a workpiece from
Buffer2).

${\bf TLLOC}_5$ for agent $\bf TU$ guarantees no overflow or underflow
of Buffer2. This is realized by counting event 4 (input a workpiece
into Buffer2) and event 5 (output a workpiece from Buffer2).

Now consider partial observation. We consider two cases: first with
$\Sigma_{uo} = \{3,6\}$, controlled behavior similar to the
full-observation case is achieved but with more complex transition
structures; second, with $\Sigma_{uo} = \{1,3,5\}$  (i.e. all
controllable events are unobservable), the resulting controlled
behavior is more restrictive.

\begin{figure}[!t]
\centering
    \includegraphics[scale = 0.8]{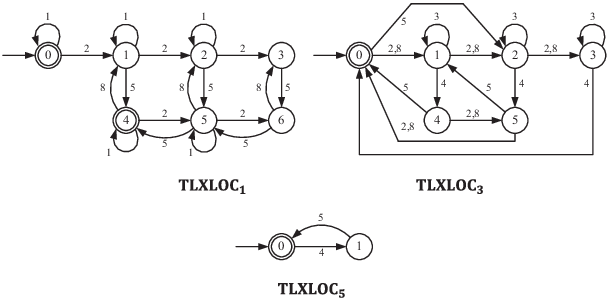}
\caption{Transfer Line: local controllers under partial observation
$P$ ($\Sigma_{uo} = \{3,6\}$)} \label{fig:TL:LOC1}
\end{figure}

Case (i) $\Sigma_{uo} = \{3,6\}$.We first compute as in
(\ref{eq:monosup}) the controllable and observable
controlled behavior ${\bf SUP1}$ which has 39 states. Then we apply
the localization algorithm to obtain the partial-observation local
controllers. The results are displayed in Fig.~\ref{fig:TL:LOC1}. It
is verified that the collective controlled behavior of these
controllers is equivalent to $\bf SUP1$.

The control logic of ${\bf TLXLOC}_1$ for agent $\bf M1$ is again to
ensure that no more than three workpieces can be processed in the
loop. But since event 6 is unobservable, the events 5 and 8
instead must be counted so as to infer the occurrences of 6: if 5
followed by 8 is observed, then 6 did not occur, but if 5 is
observed and 8 is not observed, 6 may have occurred. As can be seen
in Fig.~\ref{fig:TL:LOC1}, event 6 being unobservable increased the
structural complexity of the local controller (as compared to its
counterpart in Fig.~\ref{fig:full_obs_LOC}).

The control logic of ${\bf TLXLOC}_3$ for agent $\bf M2$ is again to
prevent overflow and underflow of the two buffers. But since event 3
is unobservable, instead the occurrences of event 4 must be observed
to infer the decrease of content in Buffer1, and at the same time
the increase of content in Buffer2. Also note that since the
unobservable controllable event 3 is enabled at states 0, 1, 2, 3, we
have selfloops of event 3 at those states. The state size of ${\bf
TLXLOC}_3$ is the same as its counterpart in
Fig.~\ref{fig:full_obs_LOC}.

${\bf TLXLOC}_5$ for agent $\bf TU$ is identical to the one in the
full-observation case.


\begin{figure}[!t]
\centering
    \includegraphics[scale = 0.6]{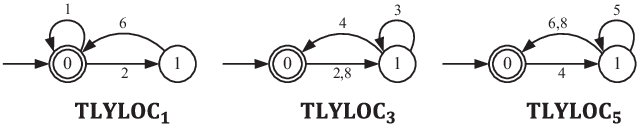}
\caption{Transfer Line: local controllers under partial observation
$P$ ($\Sigma_{uo} = \{1,3,5\}$)} \label{fig:TL:LOC2}
\end{figure}

Case (ii) $\Sigma_{uo} = \{1,3,5\}$.We first compute as in
(\ref{eq:monosup}) the controllable and observable
controlled behavior ${\bf SUP2}$ which has only 6 states. Then we
apply the localization algorithm to obtain the partial-observation
local controllers, as displayed in Fig.~\ref{fig:TL:LOC2}.

Since all the controllable events are unobservable, the controlled
behavior in this case is restrictive: ${\bf TLYLOC}_1$ for agent $\bf
M1$ allows at most one workpiece to be processed in the loop, and
${\bf TLYLOC}_2$ for agent $\bf M2$ allows at most one workpiece to be
put in Buffer1 even though Buffer1 has three slots. Also note that
in ${\bf TLYLOC}_5$ for agent $\bf TU$, since event 5 is unobservable,
events 6 and 8 instead must be observed to infer the occurrence of
5: if either 6 or 8 occurs, event 5 must have previously occurred.
In spite of the restrictive controlled behavior, these local
controllers collectively achieve equivalent controlled performance
to the 6-state $\bf SUP2$.

\begin{figure}[!t]
\centering
    \includegraphics[scale = 0.6]{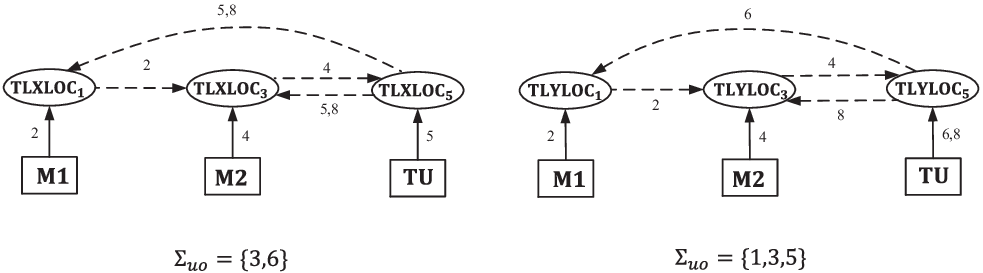}
\caption{Transfer Line: communication diagrams of local controllers.
The solid lines denote that the corresponding events are directly observed
by the local controllers; the dashed lines denote that
the corresponding events need to be communicated to the local controllers. } \label{fig:TL_com}
\end{figure}

Finally, we allocate each local controller to the agent owning the
corresponding controllable event, and according to the transition
diagrams of the local controllers, we obtain two communication diagrams one
for each case, as displayed in Fig.~\ref{fig:TL_com}. A local controller either directly
observes an event generated by the agent owning it, as denoted by the
solid lines in Fig.~\ref{fig:TL_com}, or imports an event by communication
from other local controllers, as denoted by the dashed lines.
Although the communication structures are the same in the two diagrams,
owing to different observable event sets $\Sigma_o$ the observed/communicated
events are different.


\section{Partial-Observation Localization for Large-Scale Systems} \label{sec:LargeSystem}


So far we have developed partial-observation supervisor localization
assuming that the monolithic supervisor is feasibly computable. This
assumption may no longer hold, however, when the system is
large-scale and the problem of state explosion arises.
In the literature, there have been several
architectural approaches proposed to deal with the computational
issue based on {\it model abstraction}
\cite{HillTilbury:2006,FengWonham:2008,SchmidtBreindl:2011,SuSchRoo:12}.

Just as in \cite{CaiWonham:2010a}, for large-scale system, we
propose to combine localization with an efficient heterarchical
supervisory synthesis approach \cite{FengWonham:2008} in an
alternative top-down manner: first synthesize a heterarchical array
of partial-observation decentralized supervisors and coordinators
that collectively achieves globally feasible and nonblocking
controlled behavior; then apply the developed localization algorithm
to decompose each supervisor/coordinator into local controllers for
the relevant agents.


The procedure of this heterarchical supervisor localization under
partial observation is outlined as follows:

{\it Step 1) Partial-observation decentralized supervisor
synthesis}: For each imposed control specification, collect the
relevant component agents (e.g. by event-coupling) and compute as in
(\ref{eq:SUPO}) a partial-observation decentralized
supervisor.

{\it Step 2) Subsystem decomposition and coordination}: After
Step~1, we view the system as comprised of a set of modules, each
consisting of a decentralized supervisor with its associated
component agents. We decompose the system into smaller-scale
subsystems, through grouping the modules based on their
interconnection dependencies (e.g. event-coupling or control-flow
net \cite{FengWonham:2008}).

Having obtained a set of subsystems, we verify the nonblocking
property for each of them. If a subsystem happens to be blocking, we
design a {\it coordinator} that removes blocking strings
\cite[Theorem~4]{FengWonham:2008}. The design of the coordinator
must respect partial observation; for this reason, we call the
coordinator a \emph{partial-observation coordinator}.

{\it Step 3) Subsystem model abstraction}: After Step~2, the system
consists of a set of nonblocking subsystems. Now we need to verify
the nonconflicting property among these subsystems. For this we use
model abstraction with the \emph{natural observer} property
\cite{FengWonham:2008} to obtain an abstracted model of each
subsystem.

{\it Step 4) Abstracted subsystem decomposition and coordination}:
This step is similar to Step 2, but for the abstracted models
instead of modules. We group the abstracted models based on their
interconnection dependencies, and for each group verify the
nonblocking property. If a group turns out to be blocking, we design
a partial-observation coordinator that removes blocking strings.

{\it Step 5) Higher-level abstraction}: Repeat Steps 3 and 4 until
there remains a single group of subsystem abstractions in Step 4.
The heterarchical supervisor/coordinator synthesis terminates at
Step 5; the result is a heterarchical array of partial-observation
decentralized supervisors and coordinators. Similar to
\cite{FengWonham:2008}, one can establish that these
supervisors/coordinators together achieve globally feasible and
nonblocking controlled behavior.

{\it Step 6) Partial-observation localization}: In this last step,
we apply the partial-observation localization algorithm to decompose
each of the obtained decentralized supervisors and coordinators into
local controllers for their corresponding controllable events. By
Theorem~\ref{thm:equ}, the resulting local controllers achieve the
same controlled behavior as the decentralized supervisors and
coordinators did, namely the globally feasible and nonblocking
controlled behavior.

\begin{figure}[!t]
\centering
    \includegraphics[scale = 0.20]{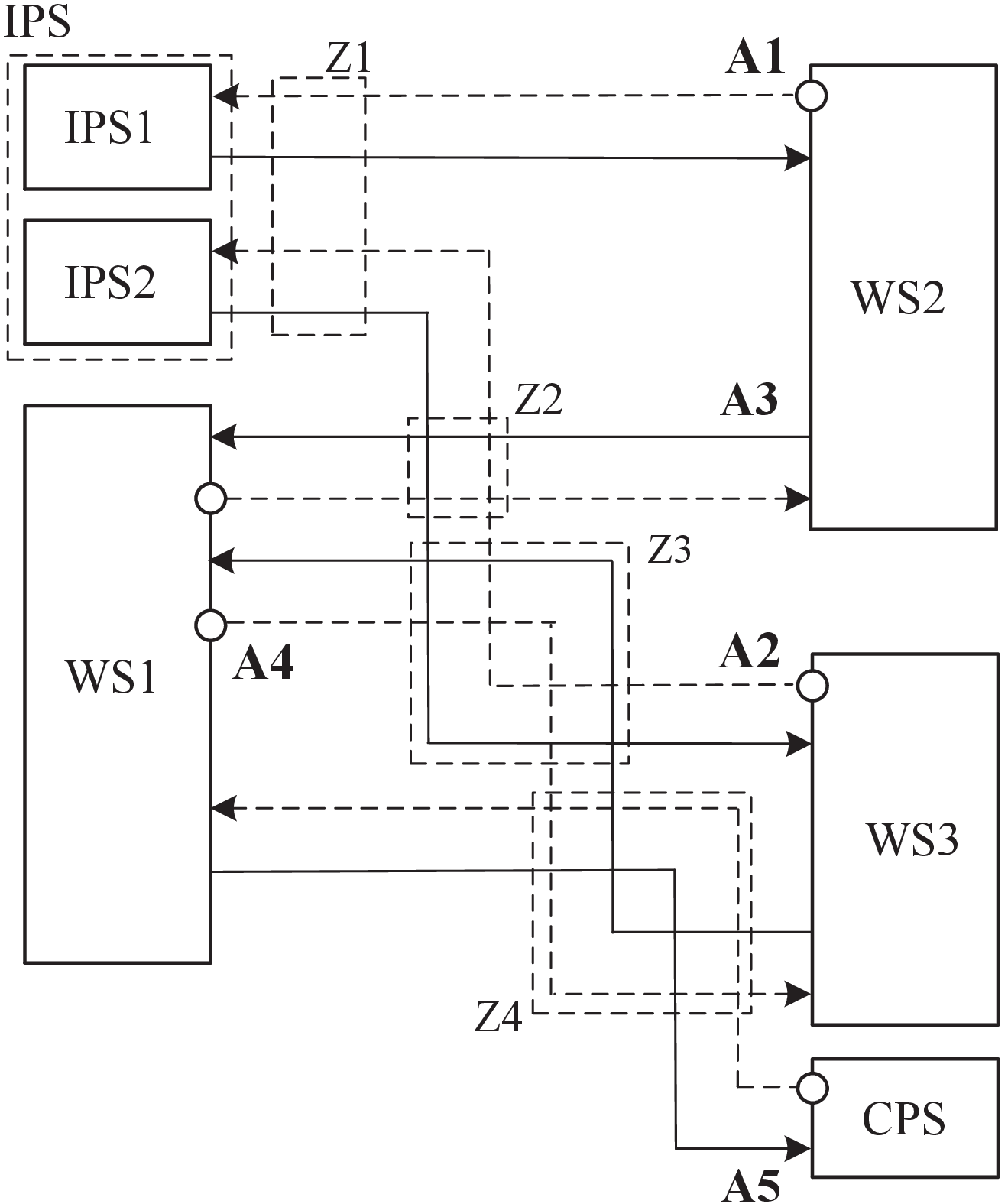}
\caption{AGV system configuration. Rectangular dashed boxes
represent shared zones of the AGV's traveling routes.}
\label{fig:AGVsystem}
\end{figure}

\begin{figure}[!t]
\centering
    \includegraphics[scale = 0.4]{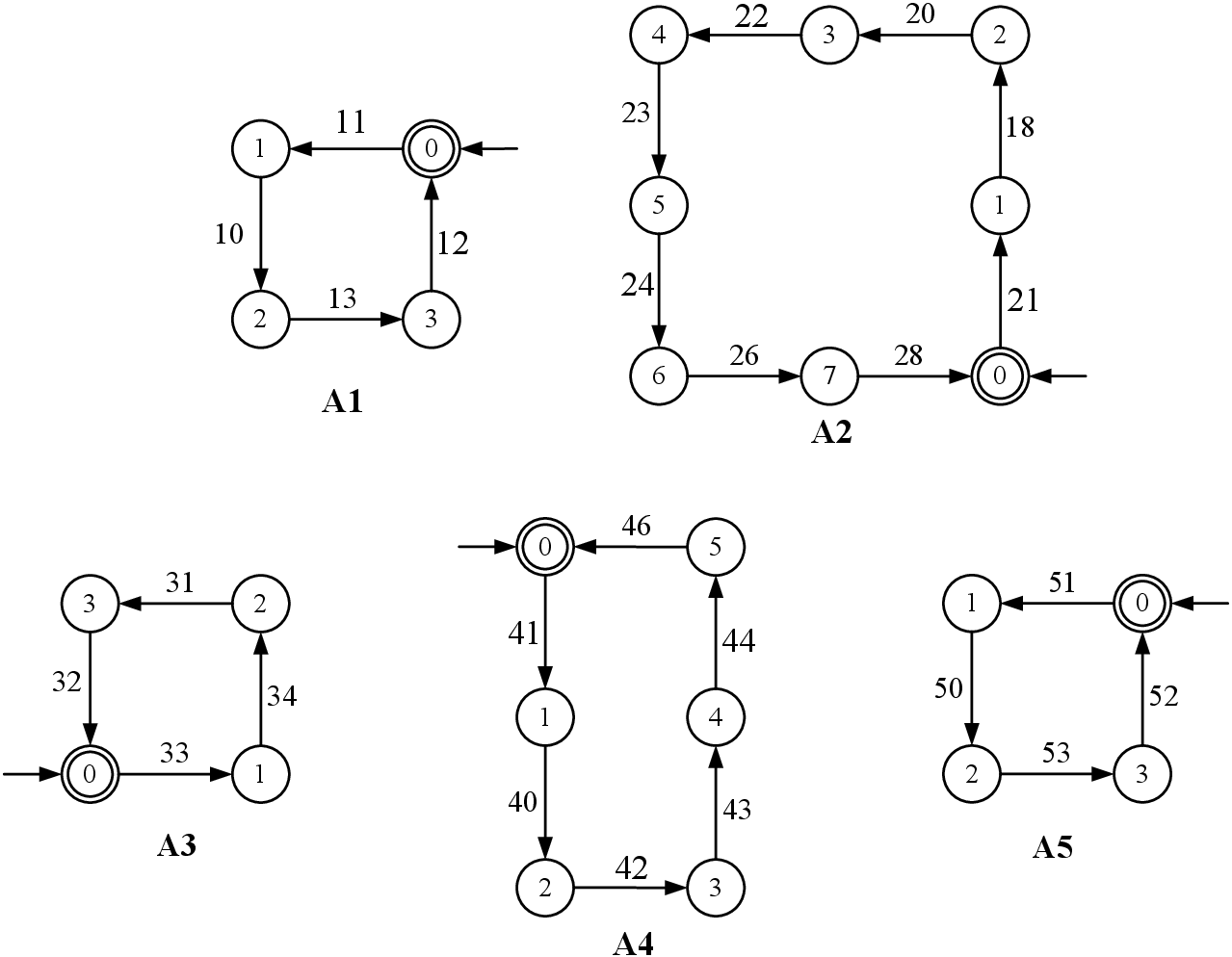}
\caption{AGV: Generators of plant components}
\label{fig:AGVplant}
\end{figure}

\begin{figure}[!t]
\centering
    \includegraphics[scale = 0.4]{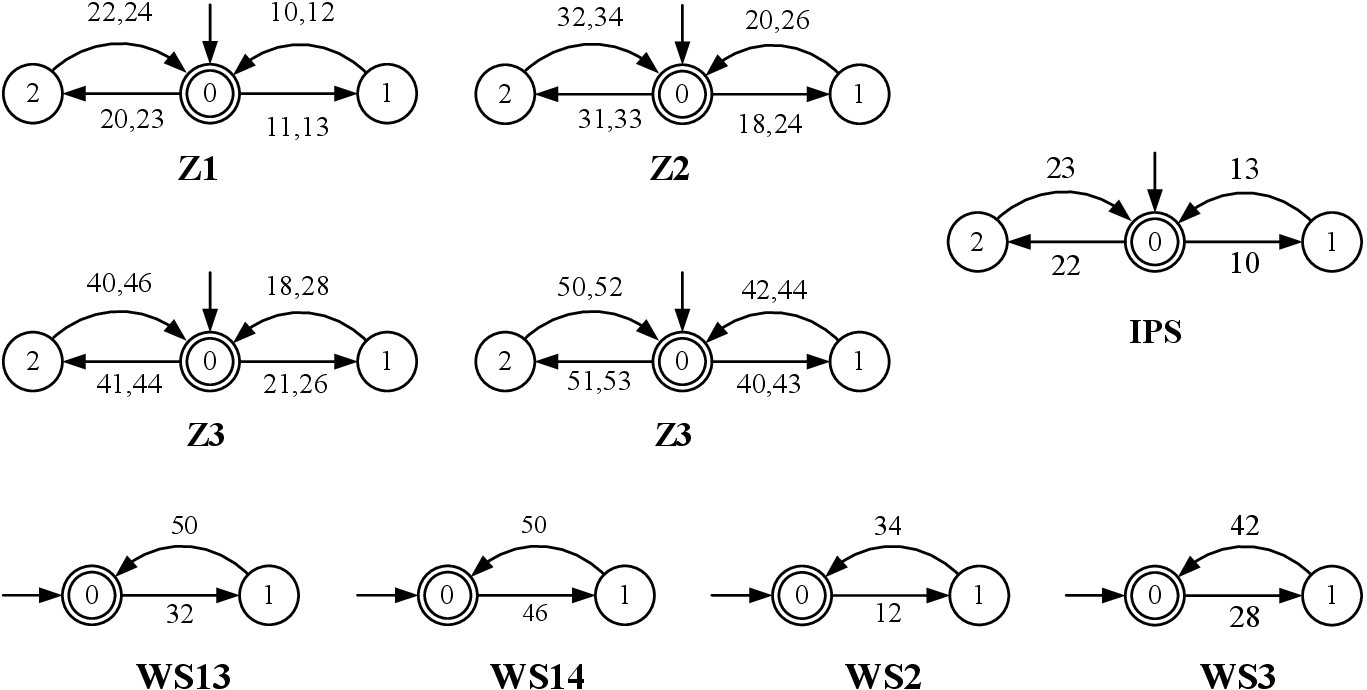}
\caption{AGV: Generators of specifications}
\label{fig:AGVspec}
\end{figure}

We note that the above procedure extends the full-observation one in
\cite{CaiWonham:2010a} by computing partial-observation
decentralized supervisors and coordinators in Steps 1-5, and finally
in Step 6 applying the partial-observation supervisor localization
developed in Section~III. In the following we apply the heterarchical localization procedure
to study the distributed control of AGV serving a manufacturing
workcell under partial observation. As displayed in
Fig.~\ref{fig:AGVsystem}, the plant consists of five independent AGV
\[{\bf A1}, {\bf A2}, {\bf A3}, {\bf A4}, {\bf A5}\]
and there are nine imposed control specifications
\begin{align*}
{\bf Z1}, {\bf Z2}, {\bf Z3}, {\bf Z3},{\bf WS13}, {\bf WS14S}, {\bf WS2}, {\bf WS3}, {\bf IPS}
\end{align*}
which require no collision of AGV in the shared zones and no
overflow or underflow of buffers in the workstations. The generator
models of the plant components and the specification are displayed in
Figs.~\ref{fig:AGVplant} and \ref{fig:AGVspec} respectively; the
detailed system description and the interpretation of the events are
referred to \cite[Section 4.7]{Wonham:2015a}.
Consider partial observation and let the unobservable event set be
$\Sigma_{uo} = \{13, 23, 31, 42, 53\}$; thus each AGV has an unobservable
event.  Our control objective is to design for each AGV a set of local strategies such that
the overall system behavior satisfies the imposed specifications and is nonblocking.

\begin{figure}[!t]
\centering
    \includegraphics[scale = 0.38]{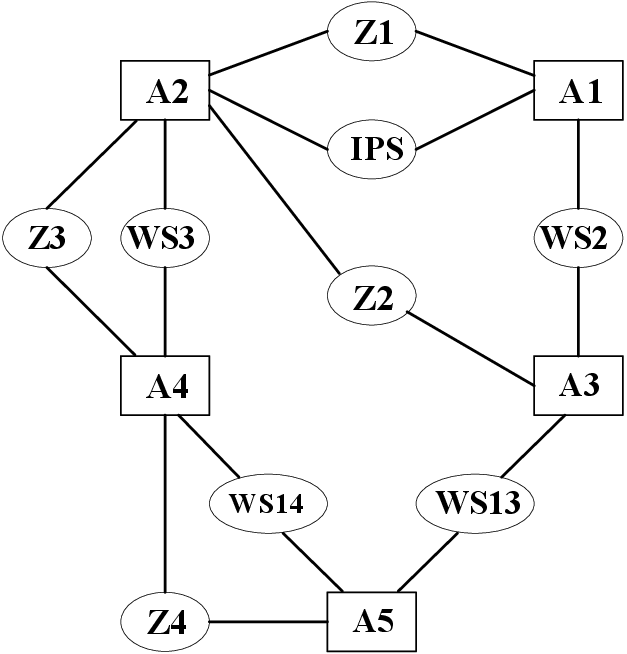}
\caption{Event-coupling relations}
\label{fig:event_couple1}
\end{figure}

{\it Step 1) Partial-observation decentralized supervisor synthesis}: For each specification
displayed in Fig.~\ref{fig:AGVspec}, we group its event-coupled AGV, as displayed in
Fig.~\ref{fig:event_couple1}, and synthesize as in (\ref{eq:SUPO}) a partial-observation
decentralized supervisor. The state sizes of these decentralized supervisors are displayed
in Table~\ref{tab:sup&coor}, in which the supervisors
are named correspondingly to the specifications, e.g. $\bf Z1SUP$
is the decentralized supervisor corresponding to the specification $\bf Z1$.

\begin{table}
\footnotesize
\caption{State sizes of partial-observation decentralized supervisors} \label{tab:sup&coor}
\begin{center}
\scalebox{0.8}{
\begin{tabular}{|c|c||c|c|}
\hline
Supervisor & State size & Supervisor & State size\\
\hline
$\bf Z1SUP$ & 13 & $\bf Z2SUP$ & 11  \\
\hline
$\bf Z3SUP$ & 26 & $\bf Z4SUP$ & 9\\
\hline
$\bf WS13SUP$ & 15 & $\bf WS14SUP$ & 19\\
\hline
$\bf WS2SUP$ & 15 & $\bf WS3SUP$ & 26\\
\hline
$\bf IPSSUP$ & 13 &  &\\
\hline
\end{tabular}
}
\end{center}
\end{table}

\begin{figure}[!t]
\centering
    \includegraphics[scale = 0.14]{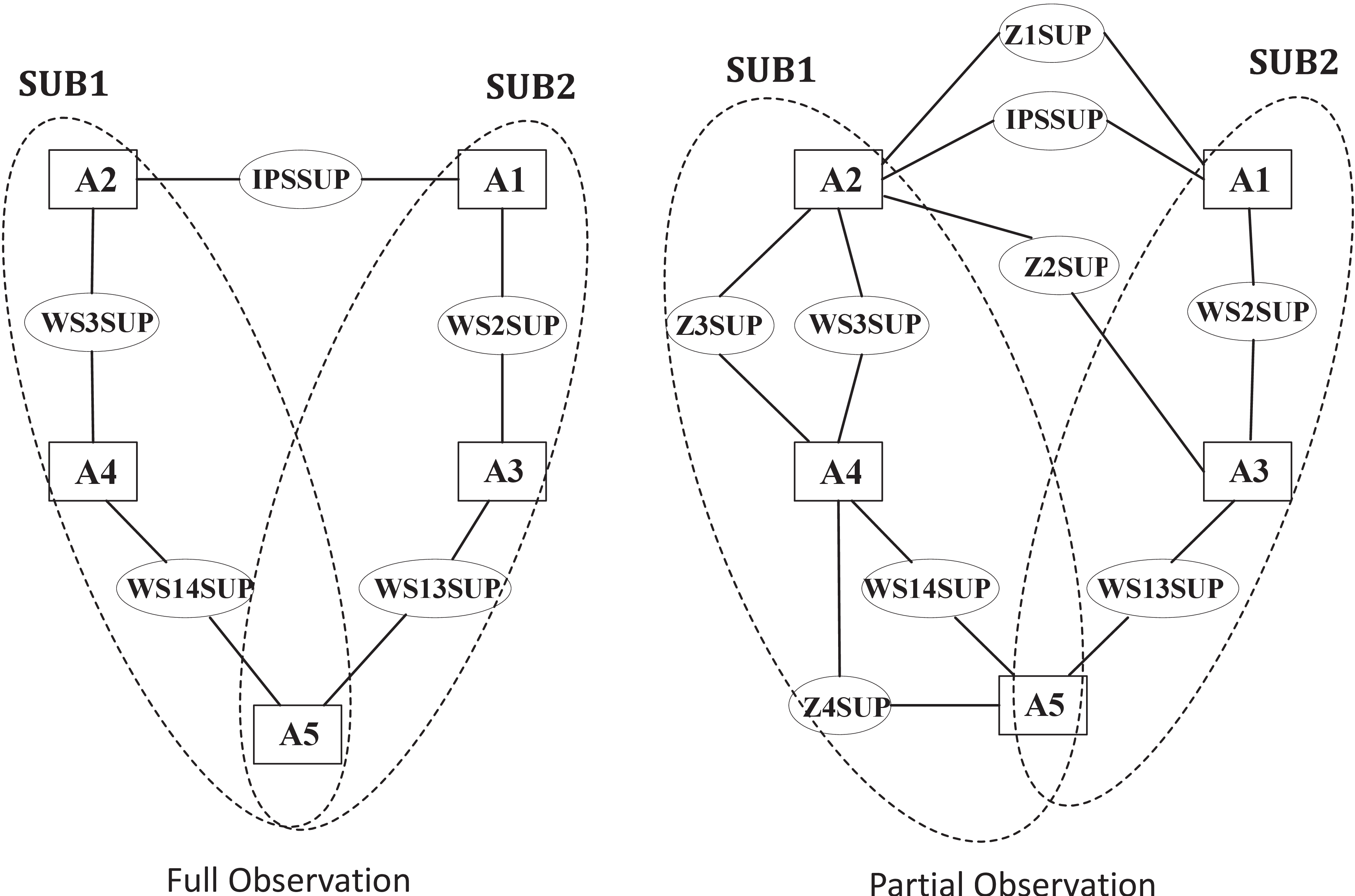}
\caption{Subsystem decomposition}
\label{fig:InterConnect}
\end{figure}

{\it Step 2) Subsystem decomposition and coordination:} We have nine
decentralized supervisors, and thus nine modules (consisting of a decentralized
supervisor with associated AGV components). Under full observation, the
decentralized supervisors for the four zones ($\bf Z1SUP$, ..., $\bf Z4SUP$)
are {\it harmless} to the overall nonblocking property \cite{FengWonham:2008},
and thus can be safely removed from the interconnection structure;
then the interconnection structure of these modules are simplified by applying
{\it control-flow net} \cite{FengWonham:2008}. Under partial observation,
however, the four decentralized supervisors are not harmless to the overall
nonblocking property and thus cannot be removed. As displayed in Fig.~\ref{fig:InterConnect},
we decompose the overall system into two subsystems:
\begin{align*}
&{\bf SUB1} := {\bf WS3SUP} || {\bf WS14SUP} || {\bf Z3SUP} || {\bf Z4SUP}\\
&{\bf SUB2} := {\bf WS2SUP} || {\bf WS13SUP}
\end{align*}
Between the two subsystems are decentralized supervisors $\bf Z1SUP$, $\bf Z2SUP$, and $\bf IPSSUP$.
It is verified that $\bf SUB2$ is nonblocking, but $\bf SUB1$ is blocking. Hence we design a
coordinator $\bf CO1$ which makes $\bf SUB1$ nonblocking, by
\[L_m({\bf CO1}) = \sup\mathcal{CO}(L_m({\bf SUB1}))\]
adapted from \cite[Theorem 4]{FengWonham:2008}.
 This coordinator $\bf CO1$ has 50 states,
and we refer to this nonblocking subsystem $\bf NSUB1$.

\begin{figure}[!t]
\centering
    \includegraphics[scale = 0.18]{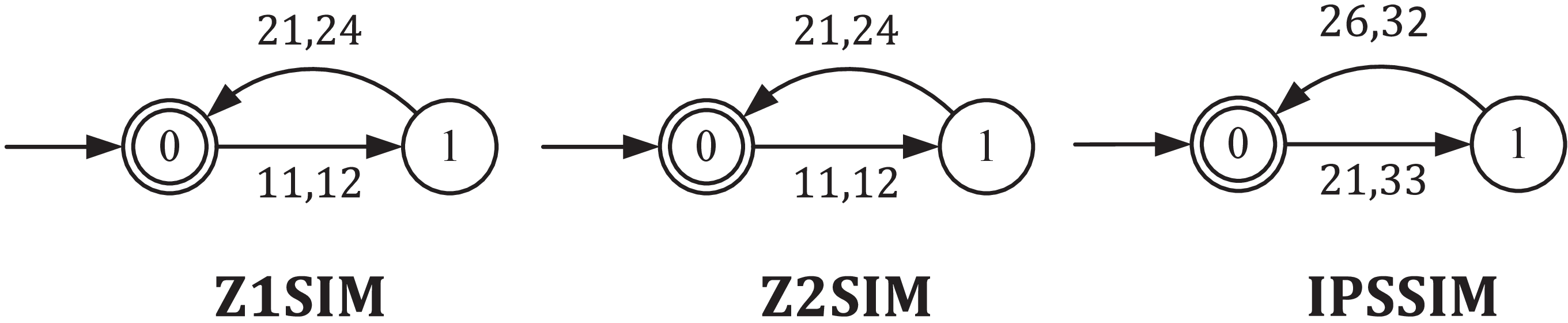}
\caption{Reduced generator models of decentralized supervisors ${\bf Z1SUP}$, ${\bf Z2SUP}$ and ${\bf IPSSUP}$}
\label{fig:simsup}
\end{figure}

{\it Step 3) Subsystem model abstraction}: Now we need to verify the
nonconflicting property among the nonblocking subsystems ${\bf NSUB1}$, ${\bf SUB2}$ and the decentralized
supervisors ${\bf IPSSUP}, {\bf Z1SUP}$ and ${\bf Z2SUP}$.
First, we determine their shared event set, denoted by $\Sigma_{sub}$.
Subsystems ${\bf NSUB1}$ and ${\bf SUB2}$ share all events in $\bf A5$:
50, 51, 52 and 53. For ${\bf IPSSUP}, {\bf Z1SUP}$ and ${\bf Z2SUP}$, we use
their reduced generator models ${\bf IPSSIM}$, ${\bf Z1SIM}$ and $\bf Z2SIM$
by supervisor reduction \cite{SuWonham:2004}, as displayed in Fig.~\ref{fig:simsup}.  By inspection, ${\bf IPSSUP}$ and ${\bf Z1SIM}$
share events 21 and 24 with ${\bf NSUB1}$, and events 11 with ${\bf SUB2}$;
${\bf Z2SUP}$ shares events 24 and 26 with ${\bf NSUB1}$, and events 32, 33 with ${\bf SUB2}$.
Thus $\Sigma_{sub} = \{11,12,21,24,26,32,33,50,51,52,53\}$.
It is then verified that $P_{sub}:\Sigma^* \rightarrow \Sigma_{sub}^*$
satisfies the natural observer property \cite{FengWonham:2008}. With $P_{sub}$, therefore,
we obtain the subsystem model abstractions, denoted by ${\bf QC\_NSUB1} = P_{sub}({\bf NSUB1})$ and
${\bf QC\_SUB2} = P_{sub}({\bf SUB2})$, with state sizes listed in Table~\ref{tab:subsys}.
\begin{table}
\footnotesize
\caption{State sizes of model abstractions} \label{tab:subsys}
\begin{center}
\scalebox{0.8}{
\begin{tabular}{|c||l|l|}
\hline
 & $\bf NSUB1$ ~~ ${\bf QC\_NSUB1}$ & $\bf SUB2$ ~~ $\bf QC\_SUB2$\\
\hline
State size & ~~~50 ~~~~~~~~~~~~~ 19 & ~~~574 ~~~~~~~~~ 56  \\
\hline
\end{tabular}
}
\end{center}
\end{table}

{\it Step 4) Abstracted subsystem decomposition and coordination}: We treat ${\bf QC\_NSUB1}$,
${\bf QC\_SUB2}$, $\bf IPSSIM$, $\bf Z1SIM$ and $\bf Z2SIM$ as a single group, and check the nonblocking property.
This group turns out to be blocking, and a coordinator $\bf CO2$ is then designed by
\begin{align*}
L_m({\bf CO2}) = \sup\mathcal{CO}(L_m({\bf QC\_SUB1})||L_m({\bf QC\_SUB2}) || \\
~~~L_m({\bf IPSSIM}) || L_m({\bf Z1SIM}) || L_m({\bf Z2SIM}))
\end{align*}
to make the group nonblocking. This coordinator $\bf CO2$ has 160 states.

%
%
%
{\it Step 5) Higher-level abstraction}: The modular supervisory control
design terminates with the previous Step 4.

%
%
We have obtained a hierarchy of nine partial-observation decentralized supervisors and
two coordinators. These supervisors and coordinators together achieve globally feasible
and nonblocking controlled behavior.

\begin{figure}[!t]
\centering
    \includegraphics[scale = 0.16]{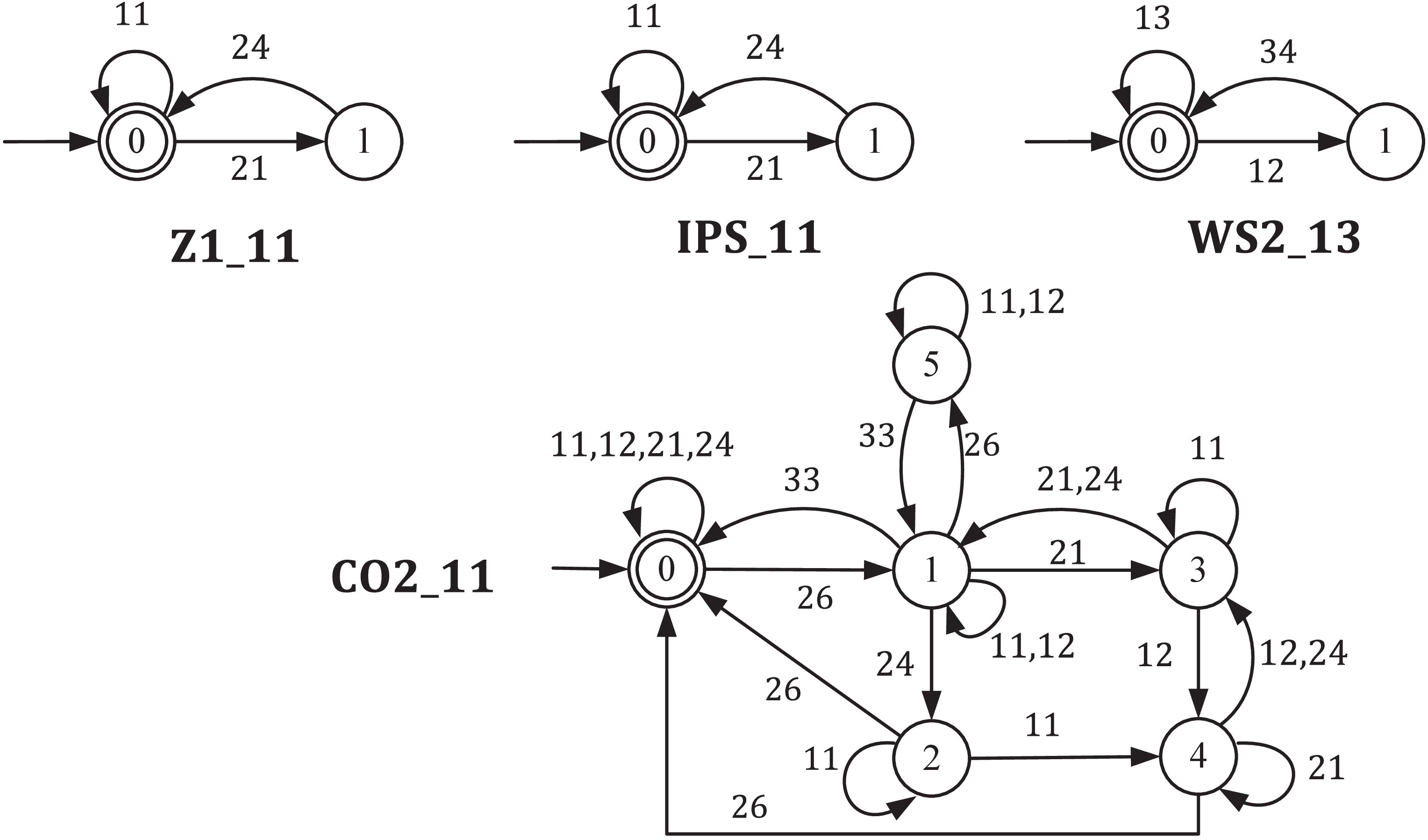}
\caption{Local controllers for $\bf A1$ with controllable events 11 and 13 (the local controllers
are named in the format of `specification\_event')}
\label{fig:LOC_A1}
\end{figure}

\begin{figure}[!t]
\centering
    \includegraphics[scale = 0.4]{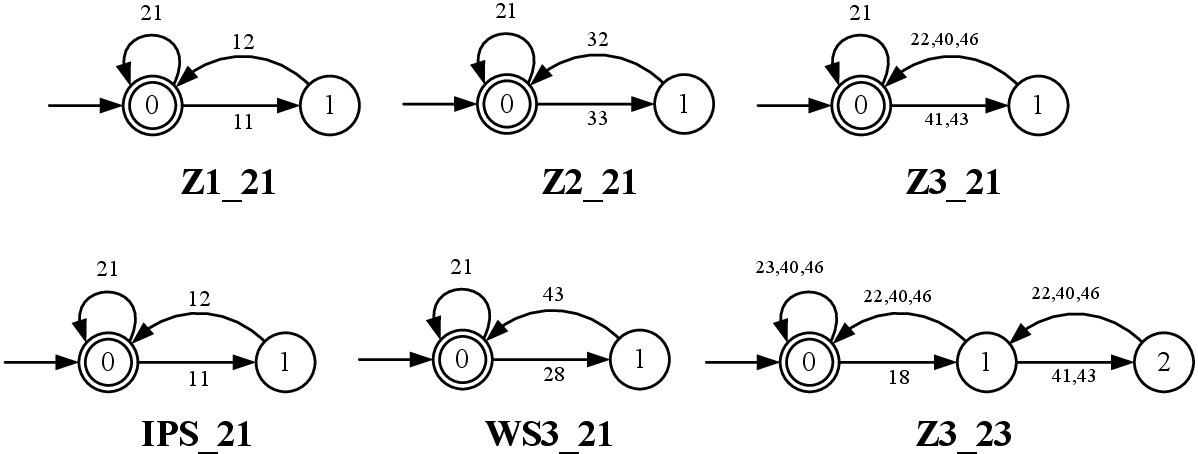}
\caption{Local controllers for $\bf A2$ with controllable events 21 and 23}
\label{fig:LOC_A2}
\end{figure}

\begin{figure}[!t]
\centering
    \includegraphics[scale = 0.16]{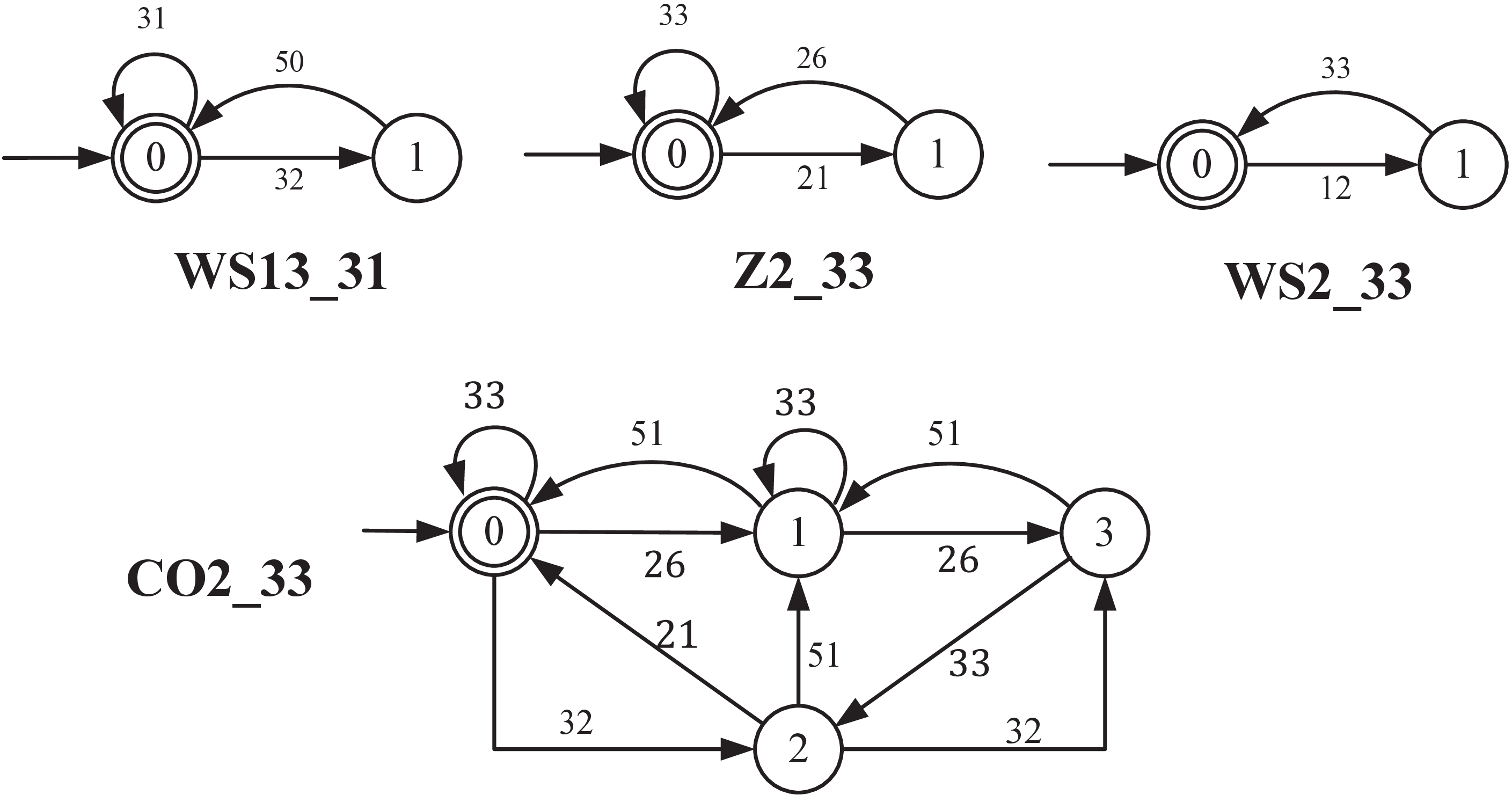}
\caption{Local controllers for $\bf A3$ with controllable events 31 and 33}
\label{fig:LOC_A3}
\end{figure}

\begin{figure}[!t]
\centering
    \includegraphics[scale = 0.14]{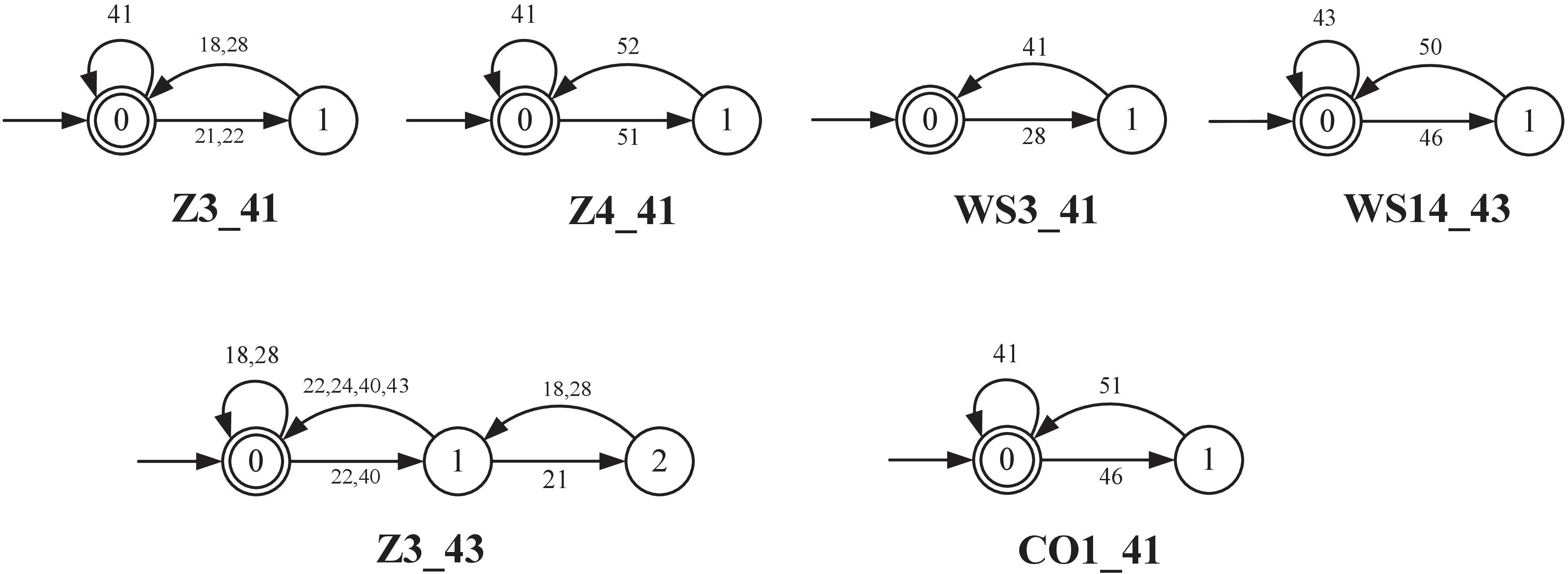}
\caption{Local controllers for $\bf A4$ with controllable events 41 and 43}
\label{fig:LOC_A4}
\end{figure}

\begin{figure}[!t]
\centering
    \includegraphics[scale = 0.4]{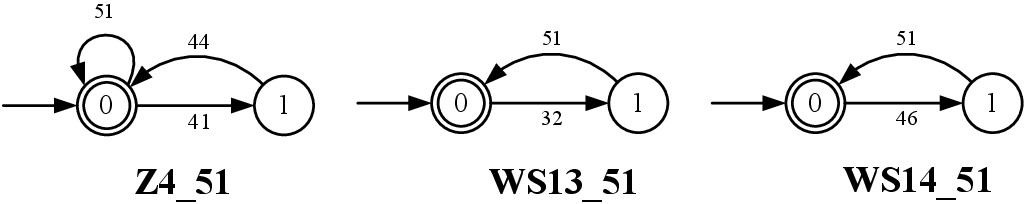}
\caption{Local controllers for $\bf A5$ with controllable events 51 and 53 (event 53 is not disabled
and thus there is no corresponding local controller)}
\label{fig:LOC_A5}
\end{figure}

{\it Step 6) Localization}: We finally apply the developed supervisor
localization procedure to decompose the obtained decentralized
supervisors/coordinators into local controllers under partial observation.
The generator models of
the local controllers are displayed in Fig.~\ref{fig:LOC_A1}-\ref{fig:LOC_A5};
they are grouped with respect to the individual AGV and their state sizes
are listed in Table~\ref{tab:loc}. By inspecting the transition structures
of the local controllers, only observable events lead to states changes.


\begin{table*}
\footnotesize
\caption{State sizes of partial-observation local controllers}
\label{tab:loc}
\begin{center}
\scalebox{0.72}{
\begin{tabular}{|c||c|c|c|c|c|}
\hline  & Local controller of & Local controller of & Local
controller of & Local controller of & Local controller of
\\
Supervisor/coordinator & $\bf A1 (state~ size)$ & $\bf A2 (state~
size)$ & $\bf A3 (state~ size)$
& $\bf A4 (state ~size)$& $\bf A5 (state ~size)$\\
\hline
$\bf Z1SUP$ & $\bf Z1\_11 (2)$ & $\bf Z1\_21 (2)$&&&\\
\hline
$\bf Z2SUP$ & &$\bf Z2\_21 (2)$&$\bf Z2\_33 (2)$&&\\
\hline
$\bf Z3SUP$ &&$\bf Z3\_21 (2)$,$\bf Z3\_23 (3)$&&$\bf Z3\_41 (2)$,$\bf Z3\_43 (3)$&\\
\hline
$\bf Z4SUP$ & &&&$\bf Z4\_41 (2)$&$\bf Z4\_51 (2)$\\
\hline
$\bf WS13SUP$ & && $\bf WS13\_31 (2)$&&$\bf WS13\_51 (2)$\\
\hline
$\bf WS14SUP$ & &&&$\bf WS14\_43 (2)$&$\bf WS14\_51 (2)$\\
\hline
$\bf WS2SUP$ & $\bf WS2\_13 (2)$&&$\bf WS2\_33 (2)$&&\\
\hline
$\bf WS3SUP$ & &$\bf WS3\_21 (2)$&&$\bf WS3\_41 (2)$&\\
\hline
$\bf IPSSUP$ & $\bf IPS\_11 (2)$&$\bf IPS\_21 (2)$&&&\\
\hline
$\bf CO1$ & &&&$\bf CO1\_41 (2)$&\\
\hline
$\bf CO2$ & $\bf CO2\_11 (6)$&&$\bf CO2\_33 (4)$&&\\
\hline
\end{tabular}
}
\end{center}
\end{table*}

Partial observation affects the control logics of the controllers
and thus affects the controlled system behavior. For illustration, consider the
following case: assuming that event sequence 11.10.13.12.21.18.20.22 has
occurred, namely $\bf A1$ has loaded a type 1 part to workstation
$\bf WS2$, and $\bf A2$ has moved to input station $\bf IPS2$. Now,
$\bf A2$ may load a type 2 part from $\bf IPS2$ (namely, event 23 may
occur). Since event 24 ($\bf A2$ exits Zone 1 and re-enter Zone 2) is
uncontrollable, to prevent the specification on Zone 2 ($\bf Z2$) not
being violated, AGV $\bf A3$ cannot enter Zone 2 if 23 has occurred,
i.e. event 33 must be disabled. However, event 33 is eligible to occur
if event 23 has occurred. So, under the full observation condition
(event 23 is observable) event 33 would occur safely if event 23 has
not occurred. However the fact is that event 23 is unobservable;
so due to (relative) observability, 33 must also be disabled even if
23 has not occurred, namely the controllers will not know whether or
not event 23 has occurred, so it will disabled event 33 in both cases,
to prevent the possible illegal behavior. This control strategy coincides
with local controller $\bf Z2\_33$: event 33 must be disabled if event
21 has occurred, and will not be re-enabled until event 26 has occurred
($\bf A2$ exits Zone 2 and re-enter Zone 3).

\begin{figure}[!t]
\centering
    \includegraphics[scale = 0.13]{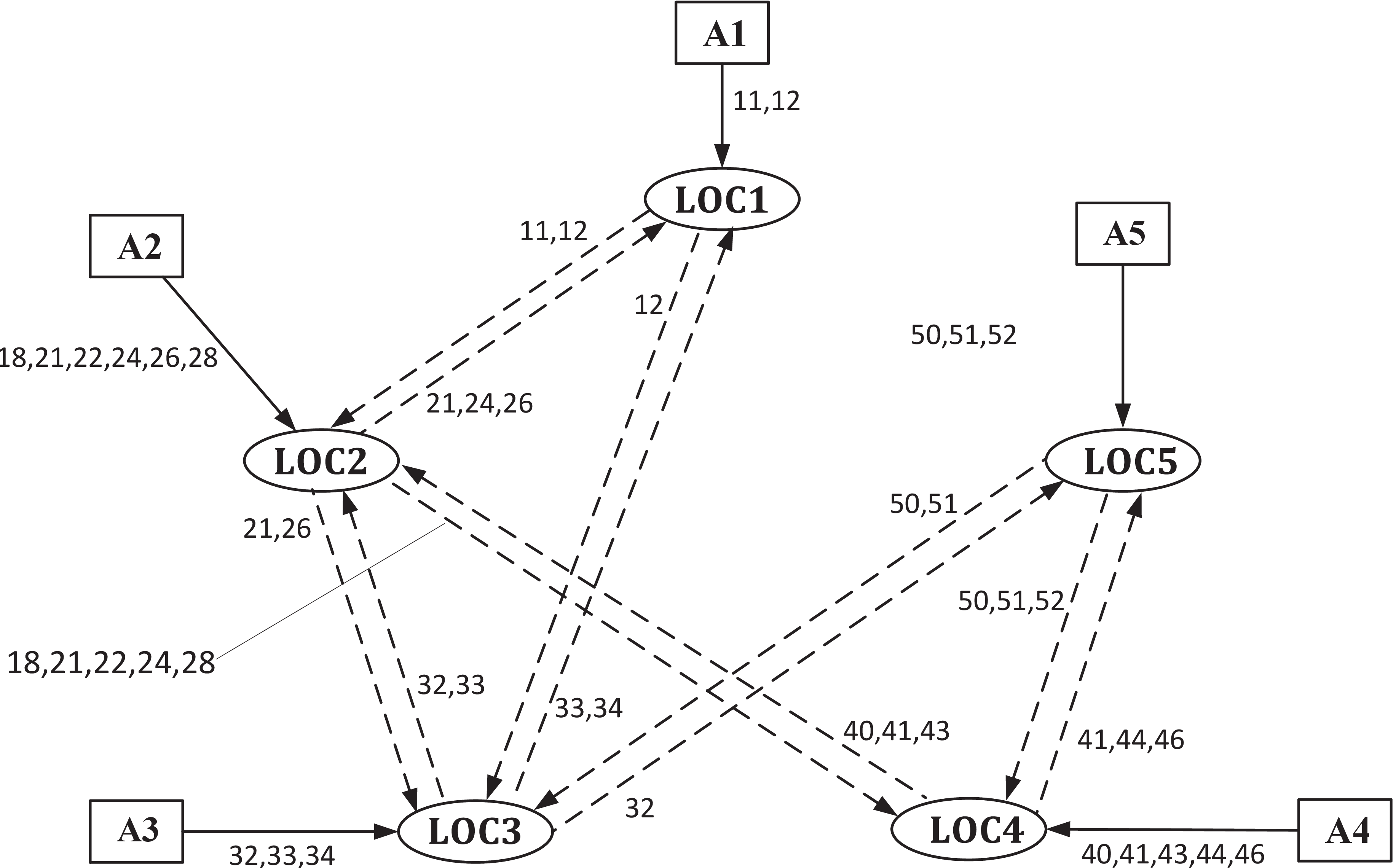}
\caption{AGV: communication diagram of local controllers. For $i = 1,...,5$,
${\bf LOCi}$ represents the local controllers corresponding to ${\bf Ai}$. } \label{fig:AGV_com}
\end{figure}

Finally, the heterarchical supervisor localization has effectively
generated a set of partial-observation local controllers with small
state sizes (between 2 and 6 states). Grouping these local
controllers for the relevant AGV, we obtain a distributed control
architecture for the system where each AGV is controlled by its own
controllers while observing certain observable events of other AGV;
according to the transition diagrams of the local controllers, we obtain a
communication diagram, as displayed in Fig.~\ref{fig:AGV_com}, which shows the events to
be observed (denoted by solid lines) or communicated (denoted
by dashed lines) to local controllers.

\section{Conclusions}
We have developed partial-observation supervisor localization to
solve the distributed control of multi-agent DES under partial
observation. This approach first employs relative observability to
compute a partial-observation monolithic supervisor, and then
decomposes the supervisor into a set of local controllers whose state
changes are caused only by observable events. A Transfer Line
example is presented for illustration. When the system is
large-scale, we have combined the partial-observation supervisor
localization with an efficient heterarchical synthesis procedure. In
future research we shall extend the partial-observation localization
procedure to study distributed control of timed DES.




\balance
\small
\bibliographystyle{IEEEtran}
\bibliography{SCDES_Ref}

\end{document}